\input harvmac
%\draftmode

\def \e {{\rm e}}
\def \ov {\over}

\def \ep {\epsilon}
\def \k {\kappa}
\def \N {{\cal N}}
\def \L {{\cal L}}
\def \K {{\cal K}}
\def \I {{\cal I}}

\def \pa { \partial}
\def \a {\alpha}
\def \E {{\cal E}}
\def \b {\beta}
\def \g {\gamma}

\def \d {\delta}

\def \l {\lambda}

\def \nn {\nabla}
\def \s {\sigma}
\def \S {\Sigma}
\def \ee {\epsilon}
\def \eps {\epsilon}
\def \r {\rho}
\def \t {\theta}

\def \p {\phi}

\def \pp {\varphi}
\def \vp {\varphi}
\def \ev {\varepsilon}

\def \ty {\tilde y}

\def \bC {\overline{C}}

\def \frac#1#2{{ #1 \over #2}}
\def \lr { \lref}

\def \td {\tilde}

\def \c {{\rm c}}
\def \M {{\cal M}}

\def \lr{\lref}

\def \gt {$G_2$\ }

\def \rf {\refs}
\def \S {\td \s}

\lr\mye{
R.~Myers,
``Superstring Gravity And Black Holes,''
Nucl.\ Phys.\ B {\bf 289}, 701 (1987).
%%CITATION = NUPHA,B289,701;%%
}

\lr\fkt{ S.~Frolov, I.~R.~Klebanov and A.~A.~Tseytlin,
``String corrections to the holographic RG flow of supersymmetric
SU(N) x SU(N+M) gauge theory,''
hep-th/0108106.
%%CITATION = HEP-TH 0108106;%%
}

\lr\howp{ P.~S.~Howe and G.~Papadopoulos,
``A Note on holonomy groups and sigma models,''
Phys.\ Lett.\ B {\bf 263}, 230 (1991).
%%CITATION = PHLTA,B263,230;%%
``Holonomy groups and W symmetries,''
Commun.\ Math.\ Phys.\  {\bf 151}, 467 (1993)
[hep-th/9202036].
%%CITATION = HEP-TH 9202036;%%
}

\lref \edel{ J.~D.~Edelstein and C.~Nunez,
``D6 branes and M-theory geometrical transitions from gauged  supergravity,''
JHEP {\bf 0104}, 028 (2001)
[hep-th/0103167].
%%CITATION = HEP-TH 0103167;%%
}

\lr\rece{
T.~Eguchi and Y.~Sugawara,
``String Theory on G2 Manifolds Based on Gepner Construction,''
hep-th/0111012.
%%CITATION = HEP-TH 0111012;%%
``CFT description of string theory compactified
on non-compact
manifolds  with G2 holonomy,''
Phys.\ Lett.\ B {\bf 519}, 149 (2001)
[hep-th/0108091].
%%CITATION = HEP-TH 0108091;%%
K.~Sugiyama and S.~Yamaguchi,
``Cascade of special holonomy manifolds and heterotic string
theory,''
hep-th/0108219.
%%CITATION = HEP-TH 0108219;%%
R.~Roiban and J.~Walcher,
``Rational Conformal Field Theories With G2 Holonomy,''
hep-th/0110302.
%%CITATION = HEP-TH 0110302;%%
R.~Blumenhagen and V.~Braun,
``Superconformal Field Theories for Compact G2 Manifolds,''
hep-th/0110232.
%%CITATION = HEP-TH 0110232;%%
}

\lr \oda{
S.~Odake,
``Extension Of N=2 Superconformal Algebra And Calabi-Yau
Compactification,''
Mod.\ Phys.\ Lett.\ A {\bf 4}, 557 (1989).
%%CITATION = MPLAE,A4,557;%%
}
\lr\figu{
J.~M.~Figueroa-O'Farrill,
``A note on the extended superconformal algebras associated with 
manifolds of exceptional holonomy,''
Phys.\ Lett.\ B {\bf 392}, 77 (1997)
[hep-th/9609113].
%%CITATION = HEP-TH 9609113;%%
}

\lr\shav{S.~L.~Shatashvili and C.~Vafa,
``Superstrings and manifold of exceptional holonomy,''
hep-th/9407025.
%%CITATION = HEP-TH 9407025;%%
}

\lr \paw {P.~S.~Howe, G.~Papadopoulos and P.~C.~West,
``Free fermions and extended conformal algebras,''
Phys.\ Lett.\ B {\bf 339}, 219 (1994)
[hep-th/9407183].
%%CITATION = HEP-TH 9407183;%%
}

\lr\cveet{
M.~Cvetic, G.~W.~Gibbons, J.~T.~Liu, H.~Lu and C.~N.~Pope,
``A new fractional D2-brane, G2 holonomy and T-duality,''
hep-th/0106162.
%%CITATION = HEP-TH 0106162;%%
M.~Cvetic, G.~W.~Gibbons, H.~Lu and C.~N.~Pope,
``M3-branes, G2 manifolds and pseudo-supersymmetry,''
hep-th/0106026.
%%CITATION = HEP-TH 0106026;%%
}

\lr\amv{ M.~Atiyah, J.~Maldacena and C.~Vafa,
``An M-theory flop as a large N duality,''
hep-th/0011256.
%%CITATION = HEP-TH 0011256;%%
}

\lref\cvet{
M.~Cvetic, G.~W.~Gibbons, H.~Lu and C.~N.~Pope,
``Cohomogeneity one manifolds of Spin(7) and G2 holonomy,''
hep-th/0108245.
%%CITATION = HEP-TH 0108245;%%
}

\lr \gres{ M.~B.~Green and S.~Sethi,
``Supersymmetry constraints on type IIB supergravity,''
Phys.\ Rev.\ D {\bf 59}, 046006 (1999)
[hep-th/9808061].
%%CITATION = HEP-TH 9808061;%%
}

\lref\PT{L.~A.~Pando Zayas and A.~A.~Tseytlin,
``3-branes on spaces with $R \times  S^2 \times  S^3$
 topology,''
Phys.\ Rev.\ D {\bf 63}, 086006 (2001)
[hep-th/0101043].
%%CITATION = HEP-TH 0101043;%%
}
\lref\PTs{L.~A.~Pando Zayas and A.~A.~Tseytlin,
``3-branes on resolved conifold,''
JHEP {\bf 0011}, 028 (2000)
[hep-th/0010088].
%%CITATION = HEP-TH 0010088;%%
}

\lref \pato{
G.~Papadopoulos and P.~K.~Townsend,
``Compactification of D = 11 supergravity on spaces of exceptional
holonomy,''
Phys.\ Lett.\ B {\bf 357}, 300 (1995)
[hep-th/9506150].
%%CITATION = HEP-TH 9506150;%%
}

\lref\bggg{
A.~Brandhuber, J.~Gomis, S.~S.~Gubser and S.~Gukov,
``Gauge theory at large N and new G2 holonomy metrics,''
Nucl.\ Phys.\ B {\bf 611}, 179 (2001)
[hep-th/0106034].
%%CITATION = HEP-TH 0106034;%%
}
\lref\bs{
R.~L.~Bryant and S.~Salamon,
``On the construction of some complete metrics with exceptional holonomy,''
Duke Math.~J. {\bf 58} (1989) 829.
}
\lref\gpp{
G.~W.~Gibbons, D.~N.~Page and C.~N.~Pope,
``Einstein Metrics On S3 R3 And R4 Bundles,''
Commun.\ Math.\ Phys.\  {\bf 127}, 529 (1990).
%%CITATION = CMPHA,127,529;%%
}

\lref \CO { P.~Candelas and X.~C.~de la Ossa,
``Comments On Conifolds,''
Nucl.\ Phys.\ B {\bf 342}, 246 (1990).
%%CITATION = NUPHA,B342,246;%%
}

\lref \cand{ P.~Candelas, M.~D.~Freeman, C.~N.~Pope, M.~F.~Sohnius and
K.~S.~Stelle,
``Higher Order Corrections To Supersymmetry And Compactifications Of
The Heterotic String,''
Phys.\ Lett.\ B {\bf 177}, 341 (1986).
%%CITATION = PHLTA,B177,341;%%
}

\lref\ts{
A.~A.~Tseytlin,
``Ambiguity In The Effective Action In String Theories,''
Phys.\ Lett.\ B {\bf 176}, 92 (1986).
%%CITATION = PHLTA,B176,92;%%
}

\lr\TTT{
A.~A.~Tseytlin,
``$R^4$ terms in 11 dimensions and
conformal anomaly of (2,0) theory,''
Nucl.\ Phys.\ B {\bf 584}, 233 (2000)
[hep-th/0005072].
%%CITATION = HEP-TH 0005072;%%
}
\lr \pvw{
K.~Peeters, P.~Vanhove and A.~Westerberg,
``Supersymmetric higher-derivative actions in ten and eleven dimensions,  the
associated superalgebras and their formulation in superspace,''
Class.\ Quant.\ Grav.\  {\bf 18}, 843 (2001)
[hep-th/0010167].
%%CITATION = HEP-TH 0010167;%%
}

\lref\PAT{ G.~Papadopoulos and A.~A.~Tseytlin,
``Complex geometry of conifolds and 5-brane wrapped on 2-sphere,''
Class.\ Quant.\ Grav.\  {\bf 18}, 1333 (2001)
[hep-th/0012034].
%%CITATION = HEP-TH 0012034;%%
}

\lref\gs{
D.~J.~Gross and J.~H.~Sloan,
``The Quartic Effective Action For The Heterotic String,''
Nucl.\ Phys.\ B {\bf 291}, 41 (1987).
%%CITATION = NUPHA,B291,41;%%
}

\lref\gz{
M.~T.~Grisaru and D.~Zanon,
``Sigma Model Superstring Corrections To
The Einstein-Hilbert Action,''
Phys.\ Lett.\ B {\bf 177}, 347 (1986).
%%CITATION = PHLTA,B177,347;%%
M.~D.~Freeman, C.~N.~Pope, M.~F.~Sohnius and K.~S.~Stelle,
``Higher Order Sigma Model Counterterms And
The Effective Action For Superstrings,''
Phys.\ Lett.\ B {\bf 178}, 199 (1986).
%%CITATION = PHLTA,B178,199;%%
}

\lr\zanon{
Q.-H.~Park and D.~Zanon,
``More On Sigma Model Beta Functions
And Low-Energy Effective Actions,''
Phys.\ Rev.\ D {\bf 35}, 4038 (1987).
%%CITATION = PHRVA,D35,4038;%%
}

\lref\grw{
D.~J.~Gross and E.~Witten,
``Superstring Modifications Of Einstein's Equations,''
Nucl.\ Phys.\ B {\bf 277}, 1 (1986).
%%CITATION = NUPHA,B277,1;%%
}

\lr\gwz{
M.~T.~Grisaru, A.~E.~van de Ven and D.~Zanon,
``Four Loop Beta Function For The N=1 And N=2
Supersymmetric Nonlinear Sigma Model In Two-Dimensions,''
Phys.\ Lett.\ B {\bf 173}, 423 (1986).
%%CITATION = PHLTA,B173,423;%%
``Two-Dimensional Supersymmetric Sigma Models
On Ricci Flat K\"ahler Manifolds Are Not Finite,''
Nucl.\ Phys.\ B {\bf 277}, 388 (1986).
%%CITATION = NUPHA,B277,388;%%
}

\lref\knw{I.~R.~Klebanov and N.~A.~Nekrasov,
``Gravity duals of fractional branes and logarithmic RG flow,''
Nucl.\ Phys.\ B {\bf 574}, 263 (2000)
[hep-th/9911096].
%%CITATION = HEP-TH 9911096;%%
I.~R.~Klebanov and E.~Witten,
``Superconformal field theory on threebranes
at a Calabi-Yau  singularity,''
Nucl.\ Phys.\ B {\bf 536}, 199 (1998)
[hep-th/9807080].
%%CITATION = HEP-TH 9807080;%%
}

\lr\pof{
M.~D.~Freeman and C.~N.~Pope,
``Beta Functions And Superstring Compactifications,''
Phys.\ Lett.\ B {\bf 174}, 48 (1986).
%%CITATION = PHLTA,B174,48;%%
}

\lr\sen{
A.~Sen,
``Central Charge Of The Virasoro Algebra
For Supersymmetric Sigma Models On Calabi-Yau Manifolds,''
Phys.\ Lett.\ B {\bf 178}, 370 (1986).
%%CITATION = PHLTA,B178,370;%%
I.~Jack and D.~R.~Jones,
``The Vanishing Of The Dilaton Beta Function For The N=2 Supersymmetric
Sigma Model,''
Phys.\ Lett.\ B {\bf 220}, 176 (1989).
%%CITATION = PHLTA,B220,176;%%
}

\lref\kts{
I.~R.~Klebanov and A.~A.~Tseytlin,
``Gravity Duals of Supersymmetric
$SU(N) \times SU(N+M)$ Gauge Theories,''
{ Nucl. Phys.} {\bf B578}, 123 (2000),
[hep-th/0002159].
%%CITATION = HEP-TH 0002159;%%
I.~R.~Klebanov and M.~J.~Strassler,
``Supergravity and a Confining Gauge Theory:
Duality Cascades and $\chi$SB-Resolution of Naked Singularities,''
JHEP {\bf 0008}, 052 (2000)
[hep-th/0007191].
%%CITATION = HEP-TH 0007191;%%
}

\lref\NS{
D.~Nemeschansky and A.~Sen,
``Conformal Invariance Of Supersymmetric Sigma Models
On Calabi-Yau Manifolds,''
Phys.\ Lett.\ B {\bf 178}, 365 (1986).
%%CITATION = PHLTA,B178,365;%%
}

\lref \MT{R.~Minasian and D.~Tsimpis,
``On the geometry of non-trivially embedded branes,''
Nucl.\ Phys.\ B {\bf 572}, 499 (2000)
[hep-th/9911042].
%%CITATION = HEP-TH 9911042;%%
K.~Ohta and T.~Yokono,
``Deformation of conifold and intersecting branes,''
JHEP {\bf 0002}, 023 (2000)
[hep-th/9912266].
%%CITATION = HEP-TH 9912266;%%
}

\lref \kaz { M.~T.~Grisaru, D.~I.~Kazakov and D.~Zanon,
``Five Loop Divergences For The N=2 Supersymmetric
Nonlinear Sigma Model,''
Nucl.\ Phys.\ B {\bf 287}, 189 (1987).
%%CITATION = NUPHA,B287,189;%%
}

\lr \MM {
J.~Maldacena,
``The large N limit of superconformal
field theories and supergravity,''
Adv.\ Theor.\ Math.\ Phys.\ {\bf 2}, 231 (1998)
[hep-th/9711200].
%%CITATION = HEP-TH 9711200;%%
S.~S.~Gubser, I.~R.~Klebanov and A.~M.~Polyakov,
``Gauge theory correlators from non-critical string theory,''
Phys.\ Lett.\ B {\bf 428}, 105 (1998)
[hep-th/9802109].
%%CITATION = HEP-TH 9802109;%%
E.~Witten,
``Anti-de Sitter space and holography,''
Adv.\ Theor.\ Math.\ Phys.\ {\bf 2}, 253 (1998)
[hep-th/9802150].
%%CITATION = HEP-TH 9802150;%%
}

\lr\gkt { S.~S.~Gubser, I.~R.~Klebanov and A.~A.~Tseytlin,
``Coupling constant dependence in the thermodynamics of N = 4
supersymmetric Yang-Mills theory,''
Nucl.\ Phys.\ B {\bf 534}, 202 (1998)
[hep-th/9805156].
%%CITATION = HEP-TH 9805156;%%
}

\lr \also{ J.~Pawelczyk and S.~Theisen,
``AdS(5) x S(5) black hole metric at O(alpha'3),''
JHEP {\bf 9809}, 010 (1998)
[hep-th/9808126].
%%CITATION = HEP-TH 9808126;%%
H.~Dorn and H.~J.~Otto,
``Q anti-Q potential from AdS-CFT relation at $ T\geq 0$: Dependence on
orientation in internal space and higher curvature corrections,''
JHEP {\bf 9809}, 021 (1998)
[hep-th/9807093].
%%CITATION = HEP-TH 9807093;%%
 K.~Landsteiner,
``String corrections to the Hawking-Page phase transition,''
Mod.\ Phys.\ Lett.\ A {\bf 14}, 379 (1999)
[hep-th/9901143].
%%CITATION = HEP-TH 9901143;%%
 T.~Harmark and N.~A.~Obers,
``Thermodynamics of spinning branes and their dual field theories,''
JHEP {\bf 0001}, 008 (2000)
[hep-th/9910036].
%%CITATION = HEP-TH 9910036;%%
``Hagedorn behaviour of little string theory from string corrections to
NS5-branes,''
Phys.\ Lett.\ B {\bf 485}, 285 (2000)
[hep-th/0005021].
%%CITATION = HEP-TH 0005021;%%
}

\lr \ndva { M.~Bertolini, P.~Di Vecchia, M.~Frau, A.~Lerda, R.~Marotta and
I.~Pesando,
``Fractional D-branes and their gauge duals,''
JHEP {\bf 0102}, 014 (2001)
[hep-th/0011077].
%%CITATION = HEP-TH 0011077;%%
J.~Polchinski,
``N = 2 gauge-gravity duals,''
Int.\ J.\ Mod.\ Phys.\ A {\bf 16}, 707 (2001)
[hep-th/0011193].
%%CITATION = HEP-TH 0011193;%%
}

\lr \PH  {
L.~Alvarez-Gaume and D.~Z.~Freedman,
``Geometrical Structure And Ultraviolet Finiteness In The Supersymmetric
Sigma Model,''
Commun.\ Math.\ Phys.\  {\bf 80}, 443 (1981).
%%CITATION = CMPHA,80,443;%%
P.~S.~Howe and G.~Papadopoulos,
``Ultraviolet Behavior Of Two-Dimensional Supersymmetric Nonlinear Sigma
Models,''
Nucl.\ Phys.\ B {\bf 289}, 264 (1987).
%%CITATION = NUPHA,B289,264;%%
}

\def \gt {$G_2$\ }

\def \CC {{\cal C}}

\lr\ppp{G. Papadopoulos, private communication.}

%%%%%%%%%%%%%%%%%%%%%%%%%%%%%%%%%%%%%%%%%%%%%%%%%%%%%
\Title{\vbox
{\baselineskip 10pt
{\hbox{OHSTPY-HEP-T-01-030}
}}}
{\vbox{\vskip -30 true pt
\centerline {$R^4$ corrections to conifolds and  $G_2$-holonomy spaces
 }
\medskip
%\centerline {~}
%\medskip
\vskip4pt }}
\vskip -20 true pt
\centerline{S. Frolov\footnote{$^*$} {Also at Steklov
Mathematical Institute, Moscow.}
 and
A.A.~Tseytlin\footnote{$^{**}$}
{Also at Imperial College, London and
Lebedev Physics Institute, Moscow.}
}
\smallskip\smallskip
\centerline{ \it  Department of Physics,
 The Ohio State University,
 Columbus, OH 43210, USA}

\bigskip\bigskip
\centerline {\bf Abstract}
\baselineskip12pt
\noindent
\medskip
Motivated by examples that appeared in the context
of string theory -- gauge theory duality,  we
consider  corrections to   supergravity
backgrounds induced  by higher derivative ($R^4$+...)
terms in superstring effective  action.
We argue that supersymmetric solutions that solve
BPS conditions
 at the leading (supergravity) order  continue to satisfy
a 1-st order ``RG-type'' system of equations
with extra  source terms encoding string (or M-theory) corrections.
We illustrate this  explicitly  on the  examples of
$R^4$ corrections to  generalized resolved and deformed
 6-d conifolds  and  a class of
non-compact  7-d  spaces with $G_2$ holonomy.
Both types of  backgrounds get non-trivial modifications
which we study in detail, stressing
 analogies between the two cases.

\bigskip

%%%%%%%%%%%%%%%%%%%%%%%%%%%%%%%%%%%%%%%%%%%%%%%%%%%%%%%%%
\Date{11/01}

%%%%%%%%%%%%%%%%%%%%%%%%%%%%%%%%%%%%%%%%%%%%%%%%%%%%%%%%%%%%%%%%%%%
\noblackbox
\baselineskip 16pt plus 2pt minus 2pt
%\baselineskip 20pt plus 2pt minus 2pt

%%%%%%%%%%%%%%%%%%%%%%%%%%%%%%%%%%%
\newsec{Introduction}
%%%%%%%%%%%%%%%%%%%
String theory -- gauge theory duality  implies certain
correspondence  between
perturbative  expansions on the both sides of the duality.
For example, in  the  case of the  AdS/CFT duality between
type IIB string theory on $AdS_5 \times S^5$  and $\N=4$ SYM theory,
the $\a'$ and $g_s$ expansions on the string side
translate into $(g^2_{\rm YM} N)^{-1/2}$ and $1/N$ expansions
on the gauge theory side \MM.
In the absence of detailed microscopic
understanding of string models  in curved Ramond-Ramond backgrounds,
one  available  source of  non-trivial  string-theoretic
information  is  the   low-energy space-time
 effective action,  which, in principle, is   universal,
 i.e.  does not depend on a  particular background.
The leading correction term in the type II string effective action
 is of  4-th power in curvature ($\a'^3 R^4$ plus additional  terms depending on other supergravity fields
as required by supersymmetry).

Provided the  basic  supergravity  background has regular (and small)
 curvature, the $\a'$-expansion is well-defined
and should contain useful
 information about strong-coupling expansion
on the gauge theory side. How the effect of $R^4$ corrections
on the string theory side translates to the gauge-theory side
was illustrated in \gkt\
on the example of the near-extremal (finite-temperature)
version of the AdS/CFT
 correspondence
(the $R^4$ correction is related to  strong-coupling
correction to  the entropy
of the $\N=4$ SYM theory, see also \also\  for some related
work).

Another example of the important role of the $R^4$ term
in the string theory -- gauge theory duality
was  given  in \fkt. There we considered the example of duality
between string theory in the ``fractional D3-brane on conifold'' background
\kts\  and $\N=1$ supersymmetric $SU(N+M) \times SU(N)$
 gauge theory  with
bifundamental matter chiral  multiplets  \knw. It was
explained how the $\a'$-corrections to
the radial dependence of  the supergravity fields
should translate into higher-order terms in the RG flow equations
on the gauge theory side. In particular, the
$R^4$ term  was  related to
the leading term in strong-coupling
expansion of  the   anomalous dimension
of matter multiplets that enters the NSVZ
 beta-functions.\foot{In contrast
to $\N=2$ examples (see, e.g., \ndva)  where the 1-loop beta-functions are exact
(and thus should be reproduced exactly  by the
dual  supergravity  backgrounds), the beta-functions of
$\N=1$
gauge theories with matter
receive  non-trivial higher-order corrections.}

The present work was motivated by  \fkt.
One general question that was left open in \fkt\
is how the $\a'$-corrected supergravity equations may be related to the RG
flow  equations on the gauge theory side
given that the former contain  higher derivatives
while the latter -- only  first order ones.
As we shall argue in section 2, in the supersymmetric cases where the
leading supergravity background  is  a solution
of 1-st order BPS system\foot{The functions $\pp^a$ parametrize
the supergravity fields
which are assumed to depend on one radial coordinate $t$ only.
This will be the case  for the examples in this paper where
the metrics  possess large global symmetry and only
the radial direction dependence is a
non-trivial one.}
 \ $\dot \pp^a = G^{ab}(\pp) {\pa W(\pp) \ov \pa\vp^b}  $ \
the solutions of the $\a'$ corrected
effective action equations should satisfy
\eqn\sati{
\dot \pp^a = \pa^a W(\pp)   + {\cal J}^a(\pp,\a')
\ ,    \ \ \ \ \ \ \   \ \ \ \   \pa^a \equiv   G^{ab}(\pp)   {\pa  \ov
\pa\vp^b} \ .  }
The ``source'' term ${\cal J}^a$ which encodes
 information about string $\a'$ corrections should
 depend only algebraically on the fields $\pp^a$.
There are two steps involved in arriving at this conclusion.
First, one believes  that,  in a supersymmetric theory,
if a  leading-order solution is supersymmetry-preserving, i.e.
solves a 1-st order system,
the same should be true for its exact $\a'$-deformation.
Indeed, one  expects that since the $\a'$-corrections
to superstring effective action ($l_P$ corrections in 11-d)
should preserve a deformed version of local supersymmetry,
the corrections to a globally-supersymmetric  background
can be found  from the deformed version of the  Killing spinor
equation  ($\nabla  \ep + \a'^3 RR \nabla R \ep +
 ...=0$). The latter
   starts with  1-st derivative term and contains
higher derivatives only  in $\a'$-dependent terms,
$ \dot \pp^a = \del^a W(\pp) +
\a'^3 B^a (\pp, \dot \pp, \ddot \pp, ...) + ...$\ .
As we shall see, one necessary condition for this to happen
is that the $R^4$-correction should vanish when
 evaluated on the leading BPS background.
Second, the fact that the leading-order background solves the
{\it first}-order
equation $\dot \pp^a = \del^a W(\pp)$  allows one to express,
order by order in $\a'$,
{\it all} derivatives of $\pp^a$ in $\a'$-correction terms  $B^a$
as  algebraic functions of $\pp^a$ only.

We shall explicitly demonstrate how  this happens
on the two examples: leading $\a'^3 R^4 $ corrections
in 10-d (or similar  $l^6_P R^4$ corrections in 11-d)
 to  (i) generalized 6-d  conifold metrics
\refs{\CO,\MT,\PTs,\PAT,\PT}
and (ii) a class of  7-d metrics with $G_2$ holonomy \refs{\bs,\gpp,\bggg}.
Both cases are Ricci flat metrics preserving  part
of global supersymmetry
(8 and 4 supercharges, respectively),
and thus  can be obtained as solutions of 1-st order systems.
These spaces have regular curvature so that
$\a'$-expansion is well-defined.
A priori,
one would expect that once $\a'$-corrections are included,
one should go back to the Einstein equations corrected
by higher-derivative terms  and solve them to find the
corrections to the
metric  (this is, in fact, what happens in bosonic string theory).
 As we shall find, in agreement with the above remarks,
 the situation  in the superstring case is
    much simpler:
the corrected solutions can be found from
   the corresponding ``corrected''
BPS equations $\dot \pp^a = \del^a  W(\pp)   + \ev
{J}^a (\pp)$, \ \  $\ev \sim \a'^3$.
 We shall  solve these equations explicitly
to leading order in $\ev $.

While the standard singular conifold metric does not receive $\a'^3$
corrections  and should be an exact string solution \fkt,
the scale-dependent  (generalized \refs{\PAT,\PT}, resolved \PTs\
and deformed \MT) conifolds get non-trivial modifications.
We shall  find that in these cases the ``source'' ${J}^a (\pp)$
is expressed in terms of a single function $E\sim RRR$ (6-d Euler density).
We shall
 relate this to the fact that the conifolds are K\"ahler manifolds,
and thus (in an appropriate scheme)
their $\a'$-deformation   may be represented  as a change  of the
K\"ahler potential \refs{\gwz,\pof,\NS}.
The $\a'$-corrections to conifold metrics we find
 provide probably the  first explicit examples
of the  expected
\refs{\gwz,\pof,\NS}  deformation of   6-d
Calabi-Yau  metrics by string $\a'$ corrections.

These results should guide the study of $\a'$-corrections
to  more general backgrounds involving p-form
fluxes on conifolds
(like the one in \refs{\kts})
which are of interest from the point of view of
 string theory -- gauge theory
duality. There the corresponding string sigma model will no longer be
K\"ahler, but the above remarks about the corrected 1-st order
form of the effective equations  will  still apply.
The $\a'$-deformation  of the metric and other fields
will then be of physical significance, being related to  higher order
corrections to RG equations on the gauge theory side
\fkt.

Below we  shall also consider the deformation
under $R^4$ corrections
of some known examples of
non-compact 7-d Ricci flat metrics
with $G_2$-holonomy \refs{\bs,\gpp,\bggg,\cvet,\cveet}
(these may also have potential gauge-theory  duality
applications as discussed, e.g.,  in \refs{\amv,\edel,\bggg}).
In the tree-level string theory context,
the corresponding NS  string sigma model has
 2-d $n=1$ world-sheet supersymmetry,  and, like in the conifold
(K\"ahler, i.e.  $n=2$) case  it is   expected
\refs{\howp,\pato}
  to be deformed  by $\a'$-corrections.\foot{Our general discussion of
 $\a'$ deformation of radially dependent Ricci flat
metrics applies also  to  simpler examples  of
hyperK\"ahler 4-d  metrics  like the Eguchi-Hanson and  Taub-NUT (or ``KK
monopole'') which correspond to
  2-d $n=4$  supersymmetric (finite \PH)
sigma model.
Here one  finds no corrections
coming from $\a'^3 R^4+...$ terms,
i.e.  in these cases there are no corrections to the
1-st order system. In general, one expects that all 1/2 supersymmetric
backgrounds of type II string theory (preserving 16 supercharges)
should not receive corrections in an appropriate scheme,
while 1/4 and 1/8 supersymmetric backgrounds should
 get corrections.
 }
We indeed find that these $G_2$ metrics are deformed
by $\a'$ corrections, implying, in particular,
a  ``deformation'' of the classical 
$G_2$-holonomy structure
which should go along with the deformation of the 
supersymmetry transformation rules and  the form of the Killing
spinor equation. At the same time, the number of
Killing spinors, i.e. the number of  corresponding 
global supersymmetries should  remain 
 the same, and that should be
reflected  in the structure of the 
associated 2-d  CFT.\foot{It would be  interesting to 
compare the
sigma model approach with the direct conformal
field theory constructions of $G_2$ CFT's in
 \refs{\shav,\paw,\figu,\rece}.
  The classical ``W-type'' symmetry of
 \gt sigma models \howp\
(associated  with covariantly constant 3-form
on target space) was promoted to a
quantum chiral algebra in \shav,
i.e.
\shav\  defined the corresponding  general
class of 2-d CFT's, with particular examples
given by particular representations of this algebra.
One expects that starting with  the quantum sigma model,
there should be a way  to define  the corresponding
quantum \gt\ algebra generators (with the definition
involving $\a'$ corrections and depending on a scheme)
so that they should 
form the same chiral quantum algebra to all
orders in $\a'$, as required by the conformal bootstrap
construction  \shav\
(we are grateful to S.  Shatashvili
for a clarifying discussion of this point).
The situation should be the same for the  6-d CY case, where
the corresponding algebra was given in \oda\
(see also \figu).
There the relation between the exact CFT and particular
2-d sigma model  may be more transparent since the 
$\a'$ corrections preserve at least  the K\"ahler structure
of the target space  metric (assuming one uses
a  renormalization  scheme where  $n=2$
world-sheet supersymmetry is manifest).
%Thus it is  only one function --
%the K\"ahler potential $K$  -- that gets deformed, and,
%moreover, one may try to argue
%that from the sigma model point of view,
%the corresponding $n=2$
%CFT is effectively defined in a non-local
%scheme \NS\ where  $K$ retains its Ricci-flat  value.
%We are grateful to G. Papadopoulos and S. Shatashvili
%for clarifying discussions of these matters.
}
% Thus it is  only one function --
%the K\"ahler potential $K$  -- that gets deformed, and, 
%moreover, one may try to argue 
%that from the sigma model point of view, 
% the corresponding $n=2$ 
%  CFT is effectively defined in a non-local 
%scheme \NS\ where  $K$ retains its Ricci-flat  value.
%%We are grateful to G. Papadopoulos and S. Shatashvili 
%for clarifying discussions of these matters. }
%if this has  implications for the  discussions
%\refs{\shav,\rece}  of the corresponding conformal field theory.
%   The classical W-type symmetry of \gt sigma models \howp\
%(associated  with covariantly constant 3-form
%on target space)
%  was  given free-field representation in  \refs{
%  \shav} (see also \paw), i.e. was translated
%into a 
%quantum chiral algebra. The question is whether there is a
%way to formulate the 2-d quantum field theory so that
% this algebra survives at the quantum level
% (it is not clear if
% there is  a non-local ``CFT scheme''
% analogous to the one in the CY case
% where its existence is implied  by a  possibility to
% deform the K\"ahler  potential of the  $n=2$ sigma model to
%  its Ricci-flat  value \NS).
% It is possible   that
% there is  a relation between the conformal anomaly and
% the anomaly
%of the  W-type  symmetry, so that the charges
%associated with the  covariantly constant
%forms may  not be conserved quantum mechanically
% \ppp.  }

The analysis of the $\a'$-corrections to the \gt metrics
has  close analogy with the simpler one in the  conifold case,
but  here we do not have the
 advantage of existence of a special scheme where
 corrections  are parametrized by a deformation of a single function --
the K\"ahler potential.
As in the CY case, the  deformed metric should
still possess the same number (four) of  global supercharges
as the leading-order one,
i.e. it should  solve the corresponding  $\a'$-deformed
version  of the  Killing spinor equation.
Again,   as in the conifold case
(and, more  generally, as
for K\"ahler Ricci flat spaces \sen),
we shall find that  there  exists
a scheme where  the dilaton is not changed from its constant
value; also,  the $R^4$ term evaluated on the leading-order solution
will vanish, 
and thus, as expected, there will be
no shift of the central charge.
 Considering \gt
spaces as
solutions of the 11-d theory, we shall find that there exists a
scheme in which the corrected solution preserves the direct-product
$R^{1,3} \times M^7$ form,
i.e. there is no generation of a warp factor.

\bigskip

The paper is organized as follows.
In section 2 we shall describe the general  approach
to deriving $\a'$-corrected form of the 1-st order
system of equations for supersymmetric (BPS) backgrounds.

In section 3 we shall  illustrate our approach on the example
of conifold metrics as solutions
of 10-d superstring theory.  We shall first
review  (in section 3.1)   the structure of  $R^4$ string tree-level
corrections to the supergravity action,
and prove that for any Ricci-flat  leading-order
solution for which $R^4$ invariant vanishes,
 there is a scheme where there is no correction
 to the dilaton. This claim  is based on  certain identity
   (proved in Appendix A)  between  second
covariant derivatives of $R^3$ invariants
(which is, in turn,   related
to the fact  that
the 4-point superstring amplitude
involving 3 gravitons and 1 dilaton vanishes).

In section 3.2  we shall  determine
the corrected form of 1-st order equations  for the generalized
resolved conifold metric and  explain that its structure
is indeed consistent with the expected \refs{\gwz,\pof,\NS}
deformation  of the K\"ahler structure. We shall
explicitly compute the corresponding
correction to the K\"ahler potential
and thus the metric of this non-compact CY space.
 In section 3.3 we shall briefly discuss
analogous computation for   the deformed conifold case.

In section 4 we shall carry out similar analysis
for a class of \gt holonomy spaces viewed as solutions
of 11-d  supergravity modified by $R^4$ terms
(similar results are found in 10-d  tree level string theory framework).
In section 4.1 we shall review the structure of the $R^4$ terms in 11-d
theory and mention correspondence upon dimensional reduction with
the string one-loop $R^4$ corrections in 10 dimensions.
In section 4.3  will  find explicitly the simple corrected form of the
BSGPP solution  \refs{\bs,\gpp}. The
analysis of the corrections to the BGGG \bggg\
metric will be  more involved and less explicit
(due to the fact that the general solution
for the homogeneous system
of 1-st order equations  describing small
 perturbations near the
BGGG solution is not known in an analytic form).

%%%%%%%%%%%%%%%%%%%%%%%%%%%%%%%%%%%%%%%%%%%%%%%%%%%%%%%%
%%%%%%%%%%%%%%%%%%%%%%%%%%%%%%%%%%%%%%%%%%%%%%%%%%%%%%%%

%%%%%%%%%%%%%%%%%%%%%%%%%%%%%%%%%%%%%%%%%%%%%%%%%%%%%%%%
\newsec{String corrections to first-order equations}
%%%%%%%%%%%%%%%%%%%%%%%%%%%%%%%%%%%%%%%%%%%%%%%%%%%%%%%%

The gravity backgrounds we shall consider
 in this
 paper will have nontrivial dependence on
 only  one ``radial" coordinate,  and may
be derived as solutions of equations of motion following
from 1-dimensional
action obtained by plugging  the ansatz for the metric
(and the dilaton)  into the
string effective action
\eqn\efaci{
S = S^{(0)} + S^{(1)}+...=\int\ dt\ \big[
{1\ov 2}G_{ab}(\pp )\dot \pp^a \dot
\pp^b\  - V(\pp ) +
\ev\L^{(1)}\big(\pp, \dot \pp, \ddot \pp, ...\big)+ ... \big]\ .}
Here
$\pp^a(t)$ are functions of the radial coordinate
that parametrize the metric and the dilaton,
$V$ is a scalar potential  following from the Einstein term in the action,
and  $\L^{(1)}$  stands for the contribution of the
leading  higher-derivative ($R^4+...$)
correction term. The expansion parameter
$\ev$ will be proportional to $\a'^3$ in $d=10$
or $l_P^6$ in $d=11$.
We will be interested in finding the corrected form of the solutions
to leading order in $\ev$  but the discussion that follows should
have a direct generalization to higher-order corrections.
The examples of spaces we shall consider will be non-singular, and thus
perturbation theory in  dimensionless ratio of $\ev$ and an
appropriate power of the curvature scale will be well-defined.

In the cases of interest, $V$ will be expressed in terms of a superpotential
(reflecting the fact that a fraction  of global
 supersymmetry  is preserved by the corresponding solutions)
\eqn\VW{
V=-{1\ov 2}G^{ab}{\pa W\ov \pa\vp^a}{\pa W\ov \pa\vp^b} \ . }
Then  the
action $S^{(0)}$ may be rewritten, up to a total derivative, as
\eqn\efac{
S^{(0)} ={1\ov 2} \int\ dt\ G_{ab}
\big(  \dot \pp^a - G^{ac}{\pa W\ov \pa\vp^c}\big)
\big( \dot \pp^b  - G^{bd}{\pa W\ov \pa\vp^d}\big)\ , }
and thus solutions of
\eqn\eqbps{
\dot \pp^a     = G^{ab}{\pa W\ov \pa\vp^c}}
provide its  extrema,
and satisfy the usual  ``zero-energy"  constraint $T+V=0$.

The string or M-theory higher-derivative corrections
to the 10-d and 11-d supergravity actions  should  preserve
a ``deformed'' version of the original local supersymmetry, so that the corrected
versions of globally-supersymmetric supergravity solutions
are expected to solve a ``deformed'' version of the Killing spinor equations,
($(\nabla_m + RR\nabla R\gamma^{...}_m+ ...) \ep=0$,
 see, e.g., \refs{\cand,\gres,\pvw}),
and thus  a  ``deformed''
version of the first-order BPS equation \eqbps.
This suggests a conjecture
(which will be justified in the  explicit examples considered below)
that the corrected
effective action \efaci\  may be rewritten in the
form similar to \efac\
\eqn\efacii{
S = {1\ov 2} \int\ dt\ G_{ab}
\big( \ \dot \pp^a - G^{ac}{\pa W\ov \pa\vp^c}
- \ev G^{ac}W_c^{(1)}
\big) \big( \ \dot \pp^b - G^{bd}{\pa W\ov \pa\vp^d}
- \ev G^{bd}W_d^{(1)}\big) + O(\ev^2) \ ,}
where $W_a^{(1)}$ are some functions of $\pp^a$'s and their  derivatives.
Then
\eqbps\  will be replaced by
\eqn\eqbpsi{
\dot \pp^a = G^{ab}{\pa W\ov \pa\vp^b} + \ev J^a \ ,\ \ \ \ \ \ \ \ \ \ \ \
J^a\equiv G^{ab}W_b^{(1)}\ .}
 $J^a$ will play the role of sources if one solves
these  corrected equations \eqbpsi\ in perturbation theory in $\ev$.
It is natural to expect that  with higher order
correction terms added to  \efaci\ and thus to \efacii,
one should be able to find the exact solutions to the resulting
effective equations from the system
\eqbpsi\ with $\ev J^a$  replaced by a
 power series in $\ev$.

Comparing \efaci\ to  \efacii, we derive the following
condition  for \efacii\ and  \eqbpsi\ to be true
(modulo total-derivative terms which we shall always ignore)
\eqn\cond{
\L^{(1)} =  - \big(  \dot \pp^a - G^{ab}{\pa
W\ov\pa\vp^b}\big)W_a^{(1)}\ .  }
One consequence  of this relation
is that the value of the  correction
$\L^{(1)}$  evaluated on the leading-order BPS solution of \eqbps\ must vanish.
As we shall see,
\cond\ will indeed be satisfied
in all examples we will consider.
We expect that this condition should
hold for any supersymmetric solution and
any correction preserving supersymmetry.
In particular,
the correction to the action
should vanish on the leading-order BPS solution.\foot{
A   heuristic reason  why this should be the case is that the value of the
correction  may  be regarded as a shift of energy
which should vanish for a supersymmetric solution.}

The condition \cond\ suggests the general  strategy of
  computing  the corrections
$J^a$ to the first-order equations \eqbpsi: one   should  introduce
the  variables \eqn\Ei{ Q^a = \dot \pp^a - G^{ab}{\pa W\ov \pa\vp^b
}\ , }
and express  the correction $\L^{(1)}$  as  a function of $\pp^a$,
$Q^a$ and  derivatives of $Q^a$.  Since  $\L^{(1)}$ should vanish
for $Q^a=0$,
expanding it  in powers of  $Q^a$ and  its derivatives we get
\eqn\gJ{
\L^{(1)} = - Q^a {\rm W}_a^{(1)}(\pp) + Q^a Q^b
 {\rm W}_{ab}^{(1)}(\pp)+\ \cdots
\ +\ {\rm total\ derivative\ terms}\ .  }
Since we are interested in
 solving  \eqbpsi\  to leading order in $\ev$, all we need to know
is the first  ${\rm W}_a^{(1)}(\pp)$ term which
we may thus identify with  ${W}_a^{(1)}$  in \cond.

In each particular case discussed below the
functions   $W_a^{(1)}(\pp)$  will be  found by a
straightforward (computer-assisted)  computation.
It is important to note that while
the correction term  $\L^{(1)}$ in \efaci\
depends,  in general,   not only on $\pp^a$ but also on its derivatives,
the leading term ${\rm W}_a^{(1)}(\pp)$ in \gJ, and,  therefore,
$J^a$, computed in this  way will depend only on $\pp^a$ --
since the leading-order equations are  {\it first}-order,
all higher derivatives of $\pp^a$ can be  expressed in terms of
$\pp^a$ using \eqbps\ order-by-order in $\ev$.

It is clear that the same should be true also at higher orders in
$\ev$, provided one uses  perturbation theory
in $\ev$. This suggests that the exact (all-order in $\ev$)
form of  the BPS equations \eqbpsi\
may admit an equivalent representation   where
{\it all} terms in the r.h.s. are simply algebraic  functions
 of $\pp^a$,
just as in  the case at the leading supergravity order in \eqbps.
One is tempted to conjecture that such ``RG equations''
which do not involve higher-derivative terms
(and resulting  after a non-trivial
rearrangement of the
$\a'$ or $l_P$ expansion) should follow
from  a   more fundamental definition of the supersymmetry/BPS
condition in string theory  which is not referring to low-energy
effective action expansion.

At the leading order
in perturbation theory in $\ev$
the corrected equations \eqbpsi\  take the form of
first-order equations.
It is useful to note  that
to compute $W_a^{(1)}(\pp)$ or  $J^a$
 we will  not need
to know the  explicit form of the  solution to the original
 equations \eqbps: after fixing a particular ansatz for the metric
 involving functions $\pp^a$ and computing the corresponding curvature invariants
 entering \efaci\ we will be able to express them in terms of $\pp^a$
 using only the general form of the first-order system \eqbps.

%%%%%%%%%%%%%%%%%%%%%%%%%%%%%%%%%%%%%%%%%%%%%%%%%%%%%%%%%%%%%%%%%
\newsec{ $R^4$ corrections to  conifold metrics
in type II superstring theory}
%%%%%%%%%%%%%%%%%%%%%%%%%%%%%%%%%%%%%%%%%%%%%%%%%%%%%%%%%%%%%%%%
In this section we will consider type II string theory on manifolds
$R^{1,3}\times M^6$,
where $M^6$ is either a
resolved or deformed conifold.
Our aim will be to determine explicitly how the metric on these spaces
is changed by the leading $R^4$ correction  to the supergravity action.
For definiteness, we will discuss the effect of the  tree-level  $\a'{}^3$ string
correction \refs{\grw,\gwz,\gz,\zanon}, but all is the same for the similar
1-loop correction (the difference between the tree-level and 1-loop
$R^4$ invariants in type IIA theory vanishes for the
6-d backgrounds we  consider).

Since the conifolds are Ricci flat  K\"ahler manifolds \CO,
one in general expects
that (in an appropriate scheme) the deformation of these  metrics will
be determined, as for all  CY metrics   \refs{\gwz,\pof,\NS},
by a modification of  the K\"ahler potential.
We shall  start  addressing this   problem  in  the  1-st order
equation  framework  of the previous section:
this approach  is more universal as it
applies to less supersymmetric (non-K\"ahler)
spaces  like $G_2$ holonomy spaces discussed later in section 4.

%%%%%%%%%%%%%%%%%%%%%%%%%%%%%%%%%%%%%%%%%%%%%%%%%%%%%%%%
\subsec{\bf  $\a'{}^3 R^4$
 terms in type II superstring effective action}
%%%%%%%%%%%%%%%%%%%%%%%%%%%%%%%%%%%%%%%%%%%%%%%%%%%%%%%%\

We begin by recalling the structure of $\a'{}^3$ corrections to the
relevant part of
type II effective action  -- the one which depends on the graviton and 
the dilaton
only.
Using the Einstein frame,
the leading $\a'{}^3$ corrections to the tree-level string
effective action implied by the structure of the Green-Schwarz
4-point massless string scattering amplitude
 can be written as \rf{\grw,\gs,\mye}
\eqn\acor{\eqalign{
 S &= {1\over 2\kappa^2 }\int\ d^{10}x  \sqrt{-g}\ [ R-{1\over
2}(\partial \p)^2  +   \ev  \L^{(1)}  ]
\ ,\ \ \ \ \ \ \ \ \ \ \ \  \L^{(1)} = \e^{-{3\ov 2}\phi} I_4(\bC )  \ ,\cr
&\ \ \
I_4(\bC ) =  \bC^{hmnk}\bC_{pmnq}\bC_h{}^{rsp}\bC^q{}_{rsk}+
{1\ov 2}\bC^{hkmn}\bC_{pqmn}\bC_h{}^{rsp}\bC^q{}_{rsk}  \ ,
}}
where $\kappa = 8 \pi^{7/2} g_s \a'^2$,
$\ev = {1\ov 8}\a'{}^3\zeta (3)$ and
\eqn\Cbar{
\overline{C}_{ijkl} = C_{ijkl} -
 {1\ov 4} \big( \nabla^2 \phi\big)_{ijkl}\ , }
\eqn\nabphi{
\big( \nabla^2 \phi\big)_{ij}{}^{kl} =
\d_i^k\nabla_j \nabla^l \phi - \d_j^k\nabla_i \nabla^l \phi -
\d_i^l\nabla_j \nabla^k \phi + \d_j^l\nabla_i \nabla^k \phi \ .}
Here $C_{ijkl}$ is the Weyl tensor, and we omit
terms proportional to
$\nabla_m \phi\nabla^m\phi$
in the definition of $\bC_{ijkl}$
(our leading-order backgrounds will have trivial dilaton field).

Due to the field redefinition ambiguity \rf{\ts,\grw},
we can assume that all the terms in \acor\ depend only on
the Weyl tensor: the Ricci tensor terms can be expressed  in terms of the dilaton
terms using leading-order equations of motion, i.e. changing the scheme).
For the same reason, we shall  also assume that  in the scheme we use
there are  no terms proportional to
the dilaton equation of motion $\nabla_m  \nabla^m\phi$.
 We will see, however, that to preserve the
K\"ahler structure of $M^6$ one  will have to change the scheme,
adding  a certain Ricci tensor dependent
term to the effective action.

Beyond the 4-point level, the above
 form of the leading  $\a'^3$ corrections is dictated by the sigma
model considerations.   As follows from \refs{\gz,\zanon},
  there exists a  scheme
in which the  metric and dilaton dependent terms in the
action (reproducing the 4-loop correction to the beta-function \gwz) are given,
in the string frame, by
\eqn\sfr{
 S= { 1 \ov 2 \k^2}
\int d^{10} x \sqrt {-g} \ e^{-2\p}
\ \big[ R  + 4(\del \phi)^2  +  \ev I_4(R)  \big] \ , }
\eqn\jnu{ I_4 (R)
= R^{hmnk} R_{pmnq} R_{h}^{\ rsp} R^{q}_{\ rsk}
\ + \ {1\over 2}  R^{hkmn} R_{pqmn} R_h^{\ rsp} R^{q}_{\ rsk}\ .
}
The action that
follows from \sfr\ upon  the
transformation $g_{mn} \to
e^{\p/2} g_{mn}$  gives the corresponding
action in  the Einstein frame that differs  from
the one in \acor\ by a change of the scheme
(assuming one restores back the presently irrelevant
 $(\del \phi)^2$ terms omitted in
\Cbar).

Since  the leading-order Ricci-flat  backgrounds we are
 interested in have
 $\phi =0$, we need to keep only
the terms  $e^{-{3\ov 2} \p} C^4$ and $C^3 \nabla^2 \phi$
in \acor\ as these
 may   give  corrections to the dilaton equation.
Explicitly, the  relevant terms in \acor\ are  found to be
\eqn\Lone{
\L^{(1)} = \e^{-{3\ov 2}\phi}\big[I_4(C) + 2E^{ij}\nabla_i\nabla_j\phi +
O((\nabla \phi)^2) \big]\ ,}
where $I_4 (C)$ is given by \jnu\ with
 $R_{ijkl}\to  C_{ijkl}$,
and $E_{ij}$ is defined by
\eqn\Eij{
E_{ij}= -C_{i}{}^{mkl}C_{jpkq}C_{l}{}^{pmq}+{1\ov
4}C_{i}{}^{mkl}C_{jmpq}C_{kl}{}^{pq}- {1\ov
2}C_{ikjl}C^{kmpq}C^{l}{}_{mpq}\ . }
The tensor $E_{ij}$  has the property
\eqn\propp{g^{ij}E_{ij}=E \ , }
where $E\sim \ep_6 \ep_6 RRR$ is the 6-d Euler density, which,
for a Ricci-flat space and
up to  a numerical   coefficient, is given  by\foot{In general, 
the 
Euler number for a (compact without boundary) manifold 
of even dimension $ d=2m$ is given by 
$
\chi_d = { 1 \ov 2^{d/2}\cdot (d/2)! \cdot  (4\pi)^{d/2} }
\int d^d x \sqrt g \ \ep_d  \ep_d   R...R $. For 
 $d=6$  we get 
$
\chi_6 =  { 1 \ov 3 \cdot 2^4\cdot (4\pi)^{3} }
\int d^6 x \sqrt g   E_6 \ , 
$ where 
$ E_6 = \ep_6  \ep_6  RRR =  64  E  +  O(R_{mn}) $, 
with    $E
= CCC + { 1 \ov 2} CCC
$ given by the expression below.
Thus 
$ \chi_6 =   { 4 \ov 3 (4\pi)^{3} }
\int d^6 x \sqrt g    E.
$}
\eqn\Eul{
E = C_{jmnk}C^{mpqn}C_{p}{}^{jk}{}_q+{1\ov 2}
C_{jkmn}C^{pqmn}C^{jk}{}_{pq}\ .}
As we will show in Appendix A, for any Einstein (in particular,
 any Ricci-flat) manifold the tensor $E_{ij}$ satisfies
 also the
following identity
\eqn\ijEij{
\nn^i\nn^j E_{ij}={1\ov 6}\nn^i\nn_i E\ .}
Thus \Lone\ may be rewritten, up to a total derivative term,
as
\eqn\Lonen{
\L^{(1)} = \e^{-{3\ov 2}\phi}\big[I_4(C) + { 1 \ov 3} E \nabla^2\phi +
O((\nabla \phi)^2) \big]\ .}
As a result,  the second term
 can be removed by the following
shift of the dilaton in \acor\
by  a local curvature invariant $E$
 \eqn\dilsh{ \p\to \p -
{1\ov 3}\ev E\ .}
 Then, in such a scheme the dilaton can get corrections
only from  the first term $\e^{-{3\ov 2}\phi}I_4(C)$ in
 \Lonen.\foot{Note that since the 4-point on-shell superstring
scattering amplitude involving 3 gravitons and 1 dilaton has the same
kinematic factor as the one following from the second term in
\Lone\ or  \Lonen, we conclude that this amplitude is always zero.}
In fact,  there  will be
{\it no } correction to the dilaton coming from the exponential
coupling of $\phi$  to the $I_4(C)$ invariant in \Lone:
for all supersymmetric backgrounds
we will  study in this  paper we will find by  direct
computation  that $I_4(C)$ {\it vanishes}.
This invariant should vanish  for all
special holonomy, supersymmetry-preserving metrics (see also \fkt)
\eqn\spee{ I_4(C)|_{\rm special\  holonomy\  Ricci\  flat\  metrics}    =0 \ . }
Thus, in the scheme where the term $E\nn^2\p$ in \Lonen\
is absent, the dilaton  of
these supersymmetric backgrounds is not modified from its
constant leading-order value.\foot{For 6-d conifolds,
these  conclusions are in agreement with
the previous general statements about the
dilaton shift (and the non-renormalization of central
charge)  for the CY spaces \refs{\sen}.}
Therefore, in  what follows we will set $\p =0$.

%%%%%%%%%%%%%%%%%%%%%%%%%%%%%%%%%%
\subsec{\bf Corrections to the resolved conifold metric}
%%%%%%%%%%%%%%%%%%%%%%%%%%%%%%%%%%%

The metric on  resolved conifold \CO\  that solves the $R_{mn}=0$ equations
is a special case of the following metric  \PTs\
\eqn\mottu{ds^2_6 =  \e^{10y} du^2 + \e^{2y} ds_5^2  \ , }
\eqn\mmmu{
ds_5^2 =  \e^{ -8w}  e_\psi^2
+  \e^{ 2w+ 2v}(e_{\theta_1}^2+ e_{\phi_1}^2)
+   \e^{ 2w - 2v} (e_{\theta_2}^2+e_{\phi_2}^2)     \ , }
where
$$
e_{\psi} = d\psi +  \cos \theta_1 d\phi_1  +  \cos \theta_2 d\phi_2  ,
\quad  e_{\theta_i}= d\theta_i\ ,  \quad  e_{\phi_i}=
\sin\theta_id\phi_i \ ,   $$
and $y,w,v$  are the functions of  the  radial coordinate $u$.
Since the leading order 10-d  background
 is the direct product $R^{1,3} \times M^6$,
it is clear from the sigma model  considerations
that there should exist a scheme where the
$\a'$-corrected  string-frame metric is also a direct product
of $R^{1,3}$ and  a corrected 6-d metric.
A priori,
%Since, in general, the dilaton
%may get corrections
% (as well as in  different scheme  choices)
 the metric need not remain a direct product in the Einstein
frame (as this property is scheme-dependent),
 so  the most general ansatz should
 be\foot{This form of the ansatz
is of course equivalent (by a redefinition of $y$ and
 $w$ in \mottu)
to  $ds^2_{10}=\e^{2 p}ds_4^2 +
ds_6^2$. }
\eqn\metcon{ ds^2_{10}=\e^{2 p}\big( ds_4^2 + ds_6^2 \big)\
,}
where  $ds_4^2$ is the 4-dimensional Minkowski` metric and $p(u)$ is an
additional ``warp factor'' function.

As was already  mentioned above,
 corrections to the metric and the dilaton can be studied separately.
Computing the scalar curvature of \metcon,\mottu, we find, as in \PTs,
 the corresponding 1-d action (${1\ov 8} \int d^{10} x\ \sqrt {- g} \ R\
 \to S^{(0)}$)
\eqn\greu{
S^{(0)}= {1\ov 2}\int du\ \e^{8p} \big[ 5 y'^2  - 5 w'^2
- v'^2 +18p'^2 + 20y'p'
+ \ { 1 \ov 4}   e^{8y} \big(
   4 e^{-2w} \cosh{ 2 v}   -  e^{-12 w} \cosh{ 4 v}  \big)
  \big] \ .
}
This action has the form \efaci\ with $\pp^a =(y,v,w,p)$
and $u$ playing  the role of the coordinate $t$.
It  admits (cf. \efac)  the following
superpotential \PTs
\eqn\suppu{
W=  - { 1 \ov 4} e^{8p+4y} ( e^{4 w}  +   e^{-6w} \cosh 2 v  )
\ . }
The corresponding system \eqbps\  of 1-st order equations is
then
\eqn\bpsu{
y' + { 1 \ov 5} e^{4y}( e^{4 w}
 +    e^{-6w} \cosh 2 v )       =0 \ ,\ \ \ \ \ \ \
w' - { 1 \ov 10} e^{4y} ( 2 e^{4w} -  3e^{-6 w} \cosh 2 v )
  =0
\ ,  }
\eqn\webu{
v' - { 1 \ov 2} e^{4y -6 w} \sinh 2 v =0 \ ,\ \ \ \  \ \ \ p'  =0 \ .}
The general solution of  these equations
depends on two non-trivial integration constants
(the third one is a shift of the radial coordinate)
and can be written as
\eqn\vvv{
\e^{-4v}=1+{6a^2\ov \r^2}\ ,\ \ \
\e^{-10w}={2\ov 3}\kappa (\r ) \e^{2v} \ ,\ \ \
\e^{2y}={1\ov 9}\kappa (\r )\r^2\e^{8w} \ , \ \ \ \  p=0  \ .  }
 The  coordinate  $\r$ and the original coordinate $u$ are related by
$
{d\r\ov du}=-\sqrt{{3\ov 2}}\e^{5y-5w-v}  ,
$
i.e. the metric \mottu\ takes the form \refs{\PTs,\PAT,\PT}
\eqn\ottu{ds^2_6 =  \kappa^{-1}(\rho) d\rho^2
+ \rho^2 \big[ { 1 \ov 9} \kappa (\rho) e_\psi^2
+ { 1 \ov 6}(  e_{\theta_1}^2+ e_{\phi_1}^2)
+ { 1 \ov 6}(1 + { 6 a^2 \ov \rho^2})(e_{\theta_2}^2+e_{\phi_2}^2) \big]
\  , }
where
\eqn\kopp{
\kappa(\r ) =  {1+{9a^2\over \r^2}-{b^6\over\r^6}\over
1+{6a^2\over \r^2} }\ , \ \ \ \ \ \ \  \r_0 \le \r <\infty\
, \ \ \ \ \ \
   \r_0^6+9a^2\r_0^4-b^6=0  \ . }
   The case of $a=b=0$ corresponds to the standard singular
   conifold
\CO; $b=0$  gives  the standard resolved conifold metric \PTs;
$a=0, b\not=0$  corresponds to the non-singular generalized
conifold
\refs{\PAT,\PT}
(a 6-d analog of the  4-d Eguchi-Hanson metric);
 $a\not=0, b\not=0$ is the generalized resolved conifold metric.

 The minimal value of $\r$, i.e. $\r_0=\r_0(a,b) \geq 0 $
is positive  and
becomes zero   when $b=0$. For $b\not=0$ we are to assume
that $\psi \in  [0, 2\pi)$ to avoid  the conical
bolt singularity at $\r=\r_0$. The  curvature invariants for this
metric
are regular (unless both $a$ and $b$ are zero when
we get back to the  singular conifold metric).

\bigskip
%%%%%%%%%%%%%%%%%%%%%%%%%%%%%%%%%%%%%%%%%%%%%%%%%%%%%%%%%%%
{\it Corrected form of 1-st order equations}
%%%%%%%%%%%%%%%%%%%%%%%%%%%%%%%%%%%%%%%%
%\bigskip

\noindent
To find the deformation of this metric  under the $R^4$ correction to the
Einstein
 action  we are to determine
 the source terms in \eqbpsi\ as  described in
section 2 (see \eqbpsi--\gJ).
Computing the $R^4$  curvature invariant  in \acor\
for the metric \mmmu\  and  then using the equations
\bpsu,\webu\  to express the derivatives of functions in terms of
functions themselves  we find  that it indeed vanishes as required
by \cond. Its first variation (cf. \gJ) gives
the sources
$J^a= ( J_y ,J_v, J_w,J_p)$   in \eqbpsi\
\eqn\sorescon{\eqalign{
J_y&= 0\ ,\ J_v=0\ , \ J_p =0\ , \cr
J_w &= \frac{8}{5}\e^{-8v - 42w - 2y}
\big( 35\e^{2v} + 73\e^{6v} + 73\e^{10v} + 35\e^{14v}
- 18\e^{10w} - 114\e^{4v + 10w} \cr
&- 168\e^{8v + 10w} -
114\e^{12v + 10w} - 18\e^{16v + 10w} + 36\e^{2v +
20w} + 111\e^{6v + 20w}\cr
& + 111\e^{10v + 20w} +
36\e^{14v + 20w} - 18\e^{4v + 30w} - 32\e^{8v +
30w} - 18\e^{12v + 30w} \big)
\ .  }}
Since the source for the $p$  equation  vanishes,
 we can thus  set it to zero $p=0$,
 i.e. the 10-d space  retains indeed its direct-product   structure.
This is a scheme-dependent property:
 if we  used another scheme, e.g.,  the one with  the Riemann
tensor $R_{ijkl}$ instead of the Weyl tensor $C_{ijkl}$ in \Lone, we would
get a nontrivial expression for the  warp factor $p$.

 The  equation for $w$  is thus  the only one that
   gets corrected,
\eqn\uhu{  w' - { 1 \ov 10} e^{4y} ( 2 e^{4w} -  3e^{-6 w} \cosh 2 v )
  = \ev J_w \ . }
 This structure of the corrected equations
 is  related to the fact that the resolved conifold is, in fact,
 a K\"ahler  manifold  (see below).
 Moreover,
 $ J_w$ can be represented as the
 $u$-derivative of the 6-d Euler
density $E$ \Eul\ evaluated on the metric \ottu\ (i.e. on
\mottu\ which solves the system
\bpsu, \webu)
\eqn\RwE{
J_w = -{4\ov 15}{d\ov du} E\ . }
%
%Indeed,
%the  Euler density $E \sim \ep_6 \ep_6 RRR$
%is given by  (for a Ricci-flat space and
%up to  a numerical   coefficient)
%\eqn\Eul{
%E = C_{jmnk}C^{mpqn}C_{p}{}^{jk}{}_q+{1\ov 2}
%C_{jkmn}C^{pqmn}C^{jk}{}_{pq}\ .}
%
Computing the 6-d Euler density  for the
metric \mottu\ and using \bpsu,\webu\ to eliminate  the derivatives
of $v,w,y$,    we get
$$
E = 6\e^{-6v - 36w - 6y}
\big( -5\e^{2v} - 8\e^{6v} - 5\e^{10v} +
3\e^{10w} +     15\e^{4v + 10w} + 15\e^{8v + 10w}~~~~~~~~~~~~~ $$
\eqn\Eulres{  + \ 3\e^{12v + 10w} - 6\e^{2v + 20w}
- 13\e^{6v + 20w} - 6\e^{10v + 20w} +
3\e^{4v + 30w} + 3\e^{8v + 30w} \big)
\ .}
Explicitly, for the metric \ottu\ we get$$
E= \frac{864}{{\rho}^{14}{( 6a^2 + {\rho}^2 ) }^7}
\big( 80a^4b^{18} + 2592a^8b^{12}{\rho}^2 +
48a^2b^{18}{\rho}^2 + 1728a^6b^{12}{\rho}^4 + 8b^{18}{\rho}^4 $$
\eqn\eed{+ \
336a^4b^{12}{\rho}^6 + 168a^4b^6{\rho}^{12} + 1296a^8{\rho}^{14} +
24a^2b^6{\rho}^{14} + 216a^6{\rho}^{16} + b^6{\rho}^{16} +
10a^4{\rho}^{18} \big)\  .
}
It follows   that $E$  vanishes
(and thus there are no corrections)
 \fkt\ for the standard (singular) conifold
which corresponds to $a=b=0$,  and is
 regular  in all other cases
 ($ E(\r_0) \geq E \geq  E(\infty)=0 $).\foot{Note that 
 since these metrics are non-compact, 
 the
 integral of \Eulres\ that appears in the formal definition of the
 Euler number (given in the footnote  above eq. \Eul)
 need   not  produce  an integer
 (we are grateful to R. Minasian and P. Vanhove 
 for drawing our
 attention to this fact).
 Indeed, 
 taking into account that the angular part of the 6-d integral 
 gives ${ 1\ov 3 \cdot 6^2} (4\pi)^3$ for $b=0$ case (where 
 $\psi$ is $4\pi$ periodic as in the standard conifold case)   
 and 
${ 1\ov 2 \cdot 3 \cdot 6^2} (4\pi)^3$
for $b\not=0$ case (where 
 $\psi$ is $2\pi$ periodic to avoid the bolt singularity)  
 and doing the integral  over $\rho$ from $\r_0$ to $\infty$
 one finds:
 $ \chi_6 (a, b=0)= 14/27 $, \ \ 
  $ \chi_6 (a, b=0)= 88/27 $.
To get an integer value  for the Euler   
number one needs to introduce a boundary at some $\r=\r_{*}$
and take into account the 
 boundary terms in the expression for the Euler number.}

\bigskip
%%%%%%%%%%%%%%%%%%%%%%%%%%%%%%%%%%%%%%%%%%%%%%%%%%%%%%%%%%%
{\it  K\"ahler structure }
%%%%%%%%%%%%%%%%%%%%%%%%%%%%%%%%%%%%%%%%
%\bigskip

\noindent
The resulting system of linear equations for $y$ in \bpsu,
$v$ in \webu\ and \uhu\  looks rather complicated, but it
 can be solved explicitly if one is  guided by
 the information provided by the existence of the
 K\"ahler structure \CO\ on  the resolved conifold.
Indeed,  the resolved conifold is a CY manifold with the
following K\"ahler potential \refs{\CO,\PTs}
\eqn\Kal{
K = 4\big[ \K (t) + a^2 \ln (1+ |\Lambda |^2 )\big] \ ,}
where $\Lambda $ is a function of angles and
 $t$ is related to the radial
coordinate $r$  used in \PTs\  as $r^2 = e^t$,
i.e. it is related to $\rho$ in \ottu\ by
\eqn\trho{
e^{2t} =  6^{-3} \left( \r^6 + 9 a^2 \r^4 -
b^6\right)\ ,\ \ \ \ \ \ \ \ \ \ -\infty\
<\ t\ <\infty \ .
}
Written in terms of
$\K$ the  resolved conifold  metric is \refs{\PTs}
\foot{The
standard form of the K\"ahler metric is
$g_{p\bar q} = \del_p \del_{\bar q} K $
where in the present case the three complex coordinates are
$U,Y, \Lambda$,  with
$U = e^{t\ov 2}  e^{{i\ov 2} (\psi + \phi_1 +\phi_2)} \cos {\theta_1
\ov2 }\cos {\theta_2\ov2 },
\ Y = e^{t\ov 2} e^{{i\ov 2} (\psi - \phi_1 +\phi_2)} \sin {\theta_1
\ov2 }\sin {\theta_2
\ov2 }$, \  $\Lambda=e^{-i\phi_2} \tan \theta_2$.}
\eqn\metrKpt{
ds_6^2= \K''\big( dt^2 + e_\psi^2\big) +
\K'(e_{\theta_1}^2+ e_{\phi_1}^2)
+ (\K'+a^2)(e_{\theta_2}^2+e_{\phi_2}^2)\ , \ \ \ \ \ \ \
\K' = { d\K\ov d t}  \ . }
It is clear from the metric that the radii of the
spheres at $t=-\infty$ (i.e. at $\r =\r_0 (a,b)$)
 are determined by $\K'(-\infty)$
which can be expressed in terms of  the constants $a$ and $b$. Comparing
\metrKpt\ with the general ansatz for the metric \mottu\ written in terms
of the $t$ coordinate, \eqn\eet{   ds^2 = e^{2y} \big[  \e^{ -8w} (dt^2 +
e_\psi^2) +  \e^{ 2w+ 2v}(e_{\theta_1}^2+ e_{\phi_1}^2)
+   \e^{ 2w - 2v} (e_{\theta_2}^2+e_{\phi_2}^2)   \big]   \ ,\ \ \ \ \
 {du\ov
dt}=\e^{-4(y+w)}\ ,  }
we get the following relations
\eqn\ywvKp{
\e^{10y}=\K'' \K'{}^2 (\K'+a^2)^2\ ,\ \ \ \ \ \  \
\e^{20w}={\K'(\K'+a^2)\ov \K''{}^2}\ , \ \ \ \ \ \e^{-4v}=1+{a^2\ov \K'}\ .}
Substituting these relations  into  \bpsu,\webu, we
find that the equations for $v$ and  $y+w$ are satisfied identically for
any function $\K$,  while the equation for
$w$ reduces to
\eqn\eqnwKp{
{1\ov 10}\big( 2t - \ln\big[\K'(\K'+a^2)\K''\big]\big) '  =0\ .}
This is, of course,  equivalent to first integral of
the only non-trivial
component of the Einstein equation
for the K\"ahler metric \metrKpt, \
$R_{m\bar n} = - \del_m
\del_{\bar n} \ln \det g = 0 , $  since the determinant of
\metrKpt\ (defined with respect to the
complex coordinates  $Y,U,\Lambda$)
 is $ e^{-4t} [\K'(\K'+a^2)\K'']^2$.

The equations \bpsu,\webu\ corrected by the sources
\sorescon,\uhu\   can be written in  the form
\eqn\bpst{
{d\ov dt}(\ty+w) + { 1 \ov 2} e^{-10w} \cosh 2 v
= 0\ ,\ \ \ \ \ \ \   {dv\ov dt} - { 1 \ov 2} e^{ -10 w} \sinh 2 v =0
\  ,  }
\eqn\forwt{
{d\ov dt}(w-{3\ov 2}\ty) - { 1 \ov 2}= -{2\ov 3}\ev {d\ov dt}E\ ,  \ \ \
 \ \ \       \ \ \ \
 \ty \equiv  y   + {4\ov 15}\ev E \ .  }
 Note that eq.  \forwt\ can be easily integrated.

The corrected 6-d metric
corresponding to the solution of this system may  be interpreted as
follows: up to a specific conformal factor (which is
related to the redefinition $y \to \td y$, cf. \eet), it is
 again
a  K\"ahler metric  with the corresponding
 K\"ahler potential satisfying
the corrected version   \forwt\     of \eqnwKp:
\eqn\eqnKpw{
 t-t_0 - {\ha} \ln\big[ \K'(\K'+a^2)\K''\big] = -{4\ov 3}\ev E  \ .}
 Here $t_0$ is an integration constant which is convenient to fix
 so that
$\e^{2t_0}={3\ov 2}$.

 Indeed,  the  Weyl transformation  $g_{mn}=e^{2 h}\td g_{mn}$
of the metric \eet, i.e. of
\mottu,\mmmu\  written in terms of the  coordinate $t$,
amounts to  the redefinition: $  y= \td y  +  h$.
Choosing $h=- {4\ov 15}\ev E$  we may  thus  express
$\td y, w, v$ in terms of the corrected K\"ahler  potential function
$\K$  using the same relations  as in \ywvKp;
then   \forwt\ reduces to \eqnKpw.

Note that $w-{3\ov 2} \ty$
is essentially the same as   $- { 1 \ov 8} \ln \det \td g$,
where $\det \td g$ is the determinant of the rescaled 6-d metric.
Eq.  \eqnKpw\ is thus  a special
case of the general
equation for the deformation of the K\"ahler potential  of a CY space
due to the string $\a'^3 R^4$ term  \refs{\pof,\NS}:
the corrected form of the beta-function equations
(in the scheme where the  K\"ahler    structure of the metric
is preserved)
is $R_{m \bar n} + \del_m \del_{\bar n}  H = 0$,
where $H$ is a series in $\a'$ of local curvature invariants,
starting with the $\a'^3 E$ term.
 That means \NS\
 that one can prepare such a K\"ahler  potential $\K
 (\a')$,
 that after all $\a'$-corrections taken into account
 one is left simply with $R_{mn}(\K_0)=- \del_m\del_{\bar n} \ln \det
 g (\K_0)  =0 $,
 where $\K_0$ corresponds to the standard CY metric,\foot{The
 two metrics -- $g(\K_0)$ and $g(\K)$ are,  of course,
 related in a complicated non-local way.}
  i.e.,
 integrating the above equation,
\eqn\rell{
\ln \det g(\K_0)  = \ln \det g(\K)    + k_1    \ev  E \ + ...
\ , \ \ \ \ \ \ \
 \ \ \ \ \ \  k_1 =  -{ 16\ov 3}  \ .  }
We shall use the  alternative  interpretation:
one starts with $\K_0$ and finds $\K= \K_0 + \ev \K_1 + ...$
as a solution of the corrected string (beta-function)
 equations.

As for the meaning of the Weyl rescaling of the metric,
it   is related to the particular scheme choice
used in \acor, which is {\it not} the same  scheme in which the metric
retains its   K\"ahler    structure (see below).

\bigskip
%%%%%%%%%%%%%%%%%%%%%%%%%%%%%%%%%%%%%%%%%%%%%%%%%%%%%%%%%%%
{\it  Corrected form of the metric  }
%%%%%%%%%%%%%%%%%%%%%%%%%%%%%%%%%%%%%%%%
%\bigskip

\noindent
Since the  curvature
 of the
 metric \ottu\ is regular (unless $a=b=0$ but then $E=0$)
 and the scale of the curvature is determined
 by either $a$ or $b$, the $\a'$-corrections  are  small
 for small enough ${\a'\ov a^2}$ or ${\a'\ov b^2}$
 (recall that $\r \ge \r_0$, see \kopp),
 the $\a'$-expansion is well-defined  and  thus it makes sense
 to concentrate on  the  leading deformation of the metric \ottu\
 caused by the first non-zero  $R^4$  correction term.

Integrating \eqnKpw\ one more time, we get (we  always expand to
leading order in $\ev$)
\eqn\eqnFp{
 \K'{}^3+{3\ov 2}a^2\K'{}^2=
 2 \int_{-\infty}^t dt'\  \e^{2t'} [1 +{8\ov 3} \ev
E(t')]  + c^3\ , }
where $c$ is another integration constant which we choose so that
 $c^3
=(\K'{}^3+{3\ov 2}a^2\K'{}^2)(-\infty)$.
Since $\K'(-\infty)$ determines the radii of
the spheres at $t=-\infty$ in the metric \eet,  such  choice of $c$ means
that we  keep
the radii unchanged by the
$R^4$ correction.
For $\ev =0$  eq. \eqnFp\ is
the equation for the K\"ahler potential of the (generalized) resolved
conifold \refs{\PTs,\PAT,\PT}: the relation to the metric \ottu,\kopp\ is
established by  setting $c= { 1 \ov 6}  b^2, \ \r^2= 6 \K'$
(see also \trho).
Eq.  \eqnFp\ implies
\eqn\olp{
\K= \K_0 + \ev \K_1 \ , \ \ \ \  \ \ \
\K'_1 =  {16\ov 9}   [(\K'_0+a^2) \K'_0]^{-1} \int_{-\infty}^t dt'\
\e^{2t'} E(t') \ . }
Changing from  $t$ to $ \rho$ coordinate
\trho\ and using \eed\ for $E$  we get
\eqn\Kcor{
 \K'_1 =  \frac{64}{{\rho}^2\,\left( 6\,a^2 + {\rho}^2 \right)
}[ \E(\r )- \E(\r_0 )]
\ ,\ \ \ \ \ \ \ \   \K_1'\equiv {d\ov dt} \K_1  \ ,  }
where $\r_0=\r_0(a,b)$ is defined in \kopp\ and
$$
\E(\r )
=\int^{t(\r)}_{+\infty} \ dt'\  \e^{2t'} E(t')
= -\frac{2}{3
{\rho}^{10}{( 6a^2 + {\rho}^2 ) }^5}
\big[ 8b^{18}{\rho}^2 + 6804a^{10}{\rho}^{10} +
5670a^8{\rho}^{12} + 3b^6{\rho}^{14} $$
\eqn\tilE{     +\
9a^4{\rho}^4\left( 42b^{12} + 42b^6{\rho}^6 + 5{\rho}^{12} \right)
+ a^2\left( 24b^{18} + 63b^6{\rho}^{12} \right)  +
54a^6( 18b^{12}{\rho}^2 + 17{\rho}^{14} ) \big] \ .}
%Differentiating $\K'$ or using \eqnKpw\ to determine $\K''$
 %we thus find the
%corrected form of the  components of the metric \metrKpt,\ywvKp.
This determines the form of the metric  \metrKpt,\ywvKp\
(which depends only on $\K'$ and $\K''$).

Note that the corrections vanish at both
ends of the interval
of values of $\r$:\  $\r_0 \leq \r < \infty$.
Eq. \Kcor\ simplifies in  the two special cases:
 the generalized conifold $a=0,\ b\not=0$, where
\eqn\eqnFpa{
 \K'_1 = {128\ov \r^4}\left(
\frac{11}{3} - \frac{b^{6}}{{\rho}^{6}} -
\frac{8 b^{18}}{3 {\rho}^{18}} \right)
\ ,\ \ \ \ \ \ \ \ \ \  b \leq \r  < \infty\ ,  }
and  the standard resolved conifold $a\not=0,\ b=0$, where
\eqn\eqnFpb{
 \K'_1 = \frac{16{\rho}^2\left( 12a^2 + {\rho}^2 \right)
    \left( 648a^4 + 126a^2{\rho}^2 + 7{\rho}^4 \right) }{3
    {( 6a^2 +  {\rho}^2 ) }^6}
\ ,\ \ \ \ \ \ \ \ \  0 \leq \r  < \infty\ .}
These expressions give
 the explicit form of the corresponding   metrics  \metrKpt.

%%%%%%%%%%%%%%%%%%%%%%%%%%%%%%%%%%%%%%%%%%%%%%%%%%%%%%%%%%%%%%
\bigskip
%%%%%%%%%%%%%%%%%%%%%%%%%%%%%%%%%%%%%%%%%%%%%%%%%%%%%%%%%%%
{\it  Scheme dependence}
%%%%%%%%%%%%%%%%%%%%%%%%%%%%%%%%%%%%%%%%
%\bigskip

\noindent
Let us now  return to the question of the Weyl rescaling of the metric
we needed to do to  make its  K\"ahler structure manifest.
This
 is related to the issue of scheme dependence.
We have shown that the $R^4$ corrections (chosen in the specific scheme \acor)
 modify the 6-dimensional metric \ottu\
into
\eqn\sdmetm{
ds_6^2 = \  \big[1 -{4\ov 15}\ev E(t)\big]\  ds_6^2(K)\ ,}
where $ds_6^2(K)$ is the K\"ahler metric defined by the K\"ahler potential
satisfying \eqnFp.
The $E$-dependent conformal factor
in front of the metric is
 scheme-dependent, i.e. it can  be  removed by a   redefinition
 of the 10-d metric. Indeed, let us
 consider  the following redefinition of the 10-d  Einstein-frame metric
 \eqn\chm{
g_{mn}\to g_{mn} +  s_1  \ev E_{mn}\ ,}
where $E_{mn}\sim RRR$ is the tensor defined in \Eij.
%$E\sim RRR$ is some combination of the cubic curvature invariants
%(not proportional to $g_{mn}$).
This redefinition changes the form of the $R^4$ correction in \acor\
by terms of the
structure $R^{mn} E_{mn} + \nabla^m \nabla^n E_{mn} + \cdots$.
Since our leading-order metric has a direct-product
$R^{1,3} \times M^6$ form,   $E_{mn}$ has non-zero components in $M_6$
directions only.
Computing $E_{mn}$  for
the resolved conifold metric \ottu, we find that in this  case\foot{Note
that this relation is of course consistent with
the general identities \propp\ and \ijEij.}
\eqn\EijE{
E_{ij} = {1\ov 6} g_{ij} E \ , \ \ \ \ \ \ \ \   i,j=1,...,6 \ . }
As a result,  the redefinition \chm\
does not change the 4-d components of the 10-d metric,
but  rescales  the 6-d components by $E$.
This implies that there is an  explicit
choice of  scheme in which
the  $R^4$-corrected  metric preserves its K\"ahler structure.
%(as was, of course, assumed in the discussions of  general CY spaces
%in  \refs{\NS}).

%%%%%%%%%%%%%%%%%%%%%%%%%%%%%%%%%%%%
\subsec{\bf Deformed conifold case }
%%%%%%%%%%%%%%%%%%%%%%%%

The  relevant  ansatz  for the
6-d   metric in this case  is  again
parametrized by  3 functions $y,w,q$  of a radial coordinate $u$
\refs{\PTs,\PAT}
   (cf. \mottu,\mmmu)
%breaking interchange  symmetry of the two $S^2$'s depends
%on 3 functions $y,w,q$  of a radial coordinate $u$
\eqn\mott{ds^2_6 =  e^{10y} du^2 + e^{2y} ds_5^2  \ , }
\eqn\mmm{
ds_5^2 =  e^{ -8w}  e_\psi^2
+  e^{ 2w+ 2q}(g_1^2+g_2^2)
+   e^{ 2w - 2q} (g_3^2+g_4^2)   \ ,  }
where  \refs{\MT}
\eqn\defr{
g_1 = - {\eps_2 +  e_{\p_1} \over\sqrt 2}\  ,\ \ \
g_2 = - {\eps_1  - e_{\t_1} \over\sqrt 2}\ , \ \ \
g_3 = {\eps_2 -  e_{\p_1}\over\sqrt 2}\  ,\  \  \
g_4 = {\eps_1 + e_{\t_1}  \over\sqrt 2}\ ,\ \ \
g_5 = e_\psi \ , }
$$
 \eps_1\equiv \sin\psi\sin\theta_2 d\phi_2+\cos\psi d\theta_2
\  , \ \ \ \ \
\eps_2
\equiv  \cos\psi\sin\theta_2 d\phi_2-\sin\psi d\theta_2\ . $$
For $q=0$ this is equivalent to  the standard conifold ansatz
or \mottu,\mmmu\ with $v=0$.
Choosing again the 10-d Einstein-frame
metric as  \metcon\ we find
that the  analog  of the Einstein action \greu\ here is
$$ S^{(0)}=
 {1\ov 2}\int du\ \e^{8p}\big[ 5 y'^2  - 5 w'^2
- q'^2 +18p'^2 + 20y'p' $$  \eqn\gre{
+ \ { 1 \ov 4}   e^{8y} \big(
   4 e^{-2w} \cosh{ 2 q}   -  e^{-12 w} -   e^{8w} \sinh^2 2q
   \big)
  \big]\ .
}
This  action  admits  the following  superpotential \PTs\
\eqn\supp{
W=   -{ 1 \ov 4} e^{8p+4y} ( e^{4 w} \cosh 2 q  +   e^{-6w}  )
\ . }
Note that this $W$  is very similar to the one in the resolved conifold case
being related to \suppu\ by  a  formal transformation
$ v  \to  q, \  e^{4 w} \leftrightarrow e^{-6w}$.
The corresponding 1-st order system \eqbps\   is
\eqn\bps{
y' + { 1 \ov 5} e^{4y}( e^{4 w}\cosh 2 q
 +    e^{-6w} ) =0 \ , \ \ \ \ \ \
w' - { 1 \ov 10} e^{4y} ( 2 e^{4w}\cosh 2 q  -  3e^{-6 w}) =0 \ ,
}
\eqn\web{
q' - { 1 \ov 2} e^{4y + 4 w} \sinh 2 q =0 \ , \ \ \ \ \ \ \
p'=0\ .}
The  solution of this system  gives the generalized
\refs{\PTs,\PAT}
deformed conifold metric
 depending on the two parameters $(b,\ep)$
and having regular curvature. For $b=0$ it becomes  the metric of
the standard
deformed conifold \refs{\CO,\MT} while for $\ep=0$ it  gives
the metric of generalized standard conifold
(i.e. \ottu,\kopp\ with $a=0$).

The system   \bps,\web\  is very similar to \bpsu,\webu\ in the
resolved conifold case
and the analysis of the $R^4$ correction to
the deformed conifold metric
is thus  closely parallel  to the one in the previous section, so
we will omit the details.

We find again that the $R^4$ invariant in \acor\ vanishes when
evaluated on the solution of \bps,\web,
and that the corrected form of the 1-st order equations  is
\eqbpsi\  with the source terms
\eqn\sordefcon{ \eqalign{
J_y= 0\ ,\ \ \ \ \  J_q=0\ , \ \ \ \ \ \ \ J_p =0\ , \ \ \ \  \
J_w = -{4\ov 15}{d\ov du} E\ .  }}
$E$ is again the 6-dimensional Euler density \Eul,
 which,  evaluated on the solution of \bps,\web,
  is given by (cf.
 \Eulres)
\eqn\Euldef{ \eqalign{
E& = \frac{3}{32}\e^{-12q - 36w - 6y}
\big( -1152\e^{12q} + 9\e^{60w} + 1152\e^{10q + 10w} +
1152\e^{14q + 10w} \cr
&- 240\e^{8q + 20w} - 1120\e^{12q +
20w} -       240\e^{16q + 20w} + 192\e^{10q + 30w} +
192\e^{14q + 30w} \cr
&-16\e^{8q + 40w} + 32\e^{12q +
40w} - 16\e^{16q + 40w} - 24\e^{2q + 50w} +
72\e^{6q + 50w} \cr
&- 48\e^{10q + 50w} - 48\e^{14q + 50w} +
72\e^{18q + 50w} - 24\e^{22q + 50w} - 22\e^{4q +
60w} \cr
& + 7\e^{8q + 60w} + 12\e^{12q + 60w} + 7\e^{16q +
60w} - 22\e^{20q + 60w} + 9\e^{24q + 60w} \big)
\ .}}
We conclude   that as in the resolved conifold case:
\noindent
  (i) this
 form  of the corrected  equations
is consistent with  the expectation  that there
 should exists a scheme
where the metric preserves its K\"ahler structure; (ii)
 in the scheme  we are using the warp factor
$p$ can be  set equal to zero, i.e. the 10-d metric
  preserves its 4+6 factorized form.

%%%%%%%%%%%%%%%%%%%%%%%%%%%%%%%%%%%%
%%%%%%%%%%%%%%%%%%%%%%%%%%%%%%%%%%%%%%%%%%%%%%%%%%%%%%%%
\newsec{$R^4$ corrections to a class of
7-d metrics with   $G_2$ holonomy}
%%%%%%%%%%%%%%%%%%%%%%%%%%%%%%%%%%%%%%%%%%%%%%%%%%%%%%%%

In this section we shall
analyze the  corrections induced by the  $R^4$ terms in the effective action
to  another class of supersymmetric
leading order  Ricci flat  solutions --
spaces with  $G_2$ holonomy found in \refs{\bs,\gpp} and in \bggg.
We shall  phrase  our discussion in the  11-d framework,
i.e. look at solutions of the $R + l^6_P R^4 + ...$ low-energy
effective  action  of M-theory,
but the  analysis of the corresponding 10-d string solutions is essentially
the same  (we shall comment on this explicitly in section 4.5).
We shall see that \gt spaces get non-trivial corrections,
implying that \gt structure should be deformed
(in line with the deformation of the local
supersymmetry transformation rule
and thus of the form of the  Killing spinor equation).

We shall derive the  corresponding corrected (``inhomogeneous'')
form  of 1-st order equations for the functions parametrizing these metrics
and discuss their solutions. The important role will be played by
the analysis of solutions of the homogeneous
part of the equations, i.e. the equations for small perturbations near
the leading-order solution. In section 4.4 we will extend (and correct)
 the analysis
of this homogeneous system given previously in \bggg.

%%%%%%%%%%%%%%%%%%%%%%%%%%%%%%%%%%%%%%%%%%%%%%%%%%%%%%%%
\subsec{\bf  $R^4$ terms in 11-dimensional theory }
%%%%%%%%%%%%%%%%%%%%%%%%%%%%%%%%%%%%%%%%%%%%%%%%%%%%%%%%

Let us start with
 recalling  the structure of the leading $R^4$  correction
 terms in the M-theory effective action (see, e.g., 
  \refs{\TTT,\pvw}
  and references there)
\eqn\twob{
S=  {1\over 2\kappa^2 }\int\ d^{11}x \sqrt{-g}\big( R
-  { 1 \ov  2 \cdot 4!} F^2_4  + \cdots \big) +  S^{(1)}  \  ,}
\eqn\acorG{\eqalign{
 S^{(1)} &= b_1 T_2 \int\ d^{11}x
 \sqrt{-g}\ ({\rm J}_0 - 2 \I_2) \ ,\cr
{\rm J}_0 &=  t_8\cdot t_8\ CCCC +
{1\ov 4} E_8 + ...\ , \ \ \ \  \
\ \ \  E_8\equiv {1\ov 3!}\ee_{11}\cdot \ee_{11} CCCC\ , \cr
\I_2 &= {1\ov 4} E_8 + 2\ee_{11}\CC_3\big[ CCCC - {1\ov 4}
(CC)^2\big] +...\ .
} }
Here $C=(C_{mnkl})$  is the Weyl tensor\foot{We choose
the scheme where the curvature tensor is expressed
in terms of the Weyl tensor and Ricci tensor terms,
with the latter replaced by the $F_4$-dependent terms  using leading-order
field equations.} and
dots in the two (super)invariants J$_0$   and $\I_2$ \ 
\TTT\
stand for
other (not completely known)
terms depending on
$F_4= d\CC_3$. Also,
\  $b_1 = (2\pi)^{-4}\cdot
3^{-2}\cdot 2^{-13}$, and
$T_2=(2\pi)^{2/3}(2\kappa^2)^{-1/3}$ (membrane tension).
The
backgrounds we will be  interested in will  have
 $F_4=0$ and the direct-product structure
$R^{1,3}\times M^7$,
where at the leading order $M^7$ will have
 $G_2$-holonomy. That means, in particular,
 that  terms depending on $F_4$ and
$\ee_{11}$ in \twob\  will not contribute,
 and the relevant part of the 11-d effective
action may be written as (cf. \acor,\sfr)
 \eqn\action{
S= S^{(0)}+ S^{(1)} ={1\over 2\kappa^2 }\int\ d^{11}x
 \sqrt{-g}\big[ R
+ \ev I_4 (C)  \big]\ ,}
\eqn\JJ{
I_4 (C)
% =3^{-1}2^{-8}J_0
= C^{hmnk}C_{pmnq}C_h{}^{rsp}C^q{}_{rsk}+
{1\ov 2}C^{hkmn}C_{pqmn}C_h{}^{rsp}C^q{}_{rsk}  \ ,
}
where  $
\ev = 3^{-1}2^{-5}(2\pi)^{-10/3}(2\kappa^2)^{2/3}$.
The direct reduction of \action\ to 10 dimensions
should give the 1-loop $R^4+...$ term in type IIA superstring
theory.
In Appendix A we perform a check of consistency
of this reduction
in the sector of $CCC\nabla \nabla \phi$ terms.

%%%%%%%%%%%%%%%%%%%%%%%%%%%%%%%%%%%%%%%%%%%%%%%%%%%%%%%%%%%%%%%%%%%%%%%%%%%
\subsec{\bf General  ansatz for the metric}
%%%%%%%%%%%%%%%%%%%%%%%%%%%%%%%%%%%%%%%%%%%%%%%%%%%%%%%%%%%%%%%%%%%
%%%%%

We shall consider a general class of  7-d
spaces  of $G_2$ holonomy studied in
\refs{\bs,\gpp,\bggg,\cvet}. They  have the global
symmetry $SU(2)\times \widetilde{SU(2)}\times U(1)\times Z_2$.
The  general ansatz for the 11-d metric
deformed by the quantum corrections that  preserves  this global
 symmetry is (cf. \mottu--\metcon)
\bggg \eqn\metric{
ds^2_{11}=e^{2p}\big( ds_4^2 + ds_7^2
\big)\ ,}
where $ds_4^2$ is the metric of the 4-dimensional Minkowski
space, and
\eqn\metrs{  \eqalign{
ds_7^2 =& \e^{4\a+4\b+2\g+2\d}dt^2 +
\e^{2\a}\big[\big(\s_1-\S_1\big)^2+ \big(\s_2-\S_2\big)^2
\big]+\e^{2\d}\big(\s_3-\S_3\big)^2
\cr &~~~~~~~~ +\
\e^{2\b}\big[\big(\s_1+\S_1\big)^2+\big(\s_2+\S_2\big)^2 \big]
+\e^{2\g}\big(\s_3+\S_3\big)^2 \ .}}
Here $\s_i$ and $\S_i$ are the basis
one-forms invariant under the left action of the
groups $SU(2)$ and $\widetilde{SU(2)}$, respectively, i.e.
\eqn\forms{
\s_1=\cos\psi d\theta + \sin\psi \sin\theta d\p\ ,\ \ \
\s_2=-\sin\psi d\theta +  \cos\psi \sin\theta d\p\ ,\ \ \
\s_3=d\psi + \cos\theta d\p\ ,}
with the analogous formulas for $\S_i$
in terms of $\td \psi, \td \theta, \td \p$.
The functions $\a,\b,\g,\d,p$ depend only on the ``radial''
coordinate $t$.

The  warp-factor field $p$
 is introduced to account for the fact that
  the $R^4$ corrections  could,  in principle,  destroy
the direct product structure of the original $R^{1,3}\times M^7$
background (this will not happen in a particular scheme we are using
but $p$ may be non-vanishing  in other schemes).

Computing the $R$-term in the 11-d action  \action\
 on this
 ansatz, one finds the following 1-d action,
  cf. \greu\  (this is the  generalization of the expression
derived
in \bggg\ to the case of nonvanishing function  $p$)
\eqn\effact{
S^{(0)} =  \int\ dt \ ( T -  V)        \ , }
$$ T =  e^{9p}( 2{\dot{\a}}^2 +
  8 \dot{\a} \dot{\b} + 2 {\dot{\b}}^2 + 4 \dot{\a} \dot{\d} +
  4\dot{\b}\dot{\d} + 4\dot{\a}\dot{\g} + 4\dot{\b}\dot{\g} +
  2\dot{\d}\dot{\g}  + 36\dot{\a}\dot{p} +
  36\dot{\b}\dot{p} + 18\dot{\d}\dot{p} +
  18\dot{\g}\dot{p} + 90{\dot{p}}^2), $$
  $$  V=
{1\ov 8} e^{9p}  \big( 2  \e^{6\a + 2\b + 2\g} -
  4 \e^{4\a + 4\b + 2\g } +
  2\e^{2\a + 6\b + 2\g } -
 8 \e^{4\a + 2\b + 2\d + 2\g} $$
  \eqn\hop{-  \ 8\e^{2\a + 4\b + 2\d + 2\g} +
  2\e^{2\a + 2\b + 4\d + 2\g} +
 \e^{4\a + 2\d + 4\g} +
  \e^{4\b + 2\d + 4\g} \big) \ . }
Comparing to \efaci, here  $\vp^a=( \a,\ \b,\ \g,\ \d,\ p)$,\ \
$\dot{\pp}^a= {d\pp^a\ov dt}$.

 This action admits
 a superpotential (in fact,
two simple superpotentials related by  interchanging
$\a\leftrightarrow\b$).
 The one that  leads to the system of first-order equations
obtained  in \bggg\ is
\eqn\Wi{
W={1\ov 2}\e^{9p}\big( 2\e^{3\a + \b + \g} + 2e^{\a + 3\b +
\g} +  2\e^{\a + \b + 2\d + \g} - \e^{2\a + \d +
2\g} + \e^{2\b + \d + 2\g} \big) \ .}
The variables and the radial coordinate used in \bggg\ are
related to $\pp^a$ and $t$ as follows
\eqn\ABCD{
A=\e^\a\ ,\ \ \ B=\e^\b \ , \ \ \ C=\e^\g\ ,\ \ \ D=\e^\d\ ,\ \ \
dr=\e^{2\a+2\b+2\g+\d} dt \ ,}
\eqn\meet{  \eqalign{
ds_7^2 =&  C^{-2} (r) dr^2 +A^2(r)
\big[\big(\s_1-\S_1\big)^2+ \big(\s_2-\S_2\big)^2
\big]+  D^2(r)  \big(\s_3-\S_3\big)^2
\cr &~~~~~~~~ +\
B^2(r) \big[\big(\s_1+\S_1\big)^2+\big(\s_2+\S_2\big)^2 \big]
+ C^2(r)  \big(\s_3+\S_3\big)^2 \ .}}
The superpotential \Wi\  expressed in terms of $A,B,C,D$ takes the form
\eqn\WiA{
W= \e^{9p}\big(  A^3BC + AB^3C + ABCD^2 - {1\ov 2}A^2C^2D + {1\ov 2}B^2C^2D
\big) \ .}
The corresponding system of
first-order equations \eqbps\ is the one
studied in \bggg\
  $$
{dA\ov dr}= {1\ov 4}\big({B^2-A^2+D^2\ov BCD}+{1\ov A}\big)
\ , \ \ \ \ \ \
{dB\ov dr} = {1\ov 4}\big({A^2-B^2+D^2\ov ACD}-{1\ov B}\big)\  , $$
\eqn\system{
{dC\ov dr}= {1\ov 4}  \big( {C\ov B^2}-{C\ov A^2}\big)\ , \ \ \ \ \
{dD\ov dr}=  {A^2+B^2-D^2\ov 2ABC} \ , \ \ \ \ \ \
{dp\ov dr}= 0 \ .
}
The superpotential given  in \bggg\foot{Note  that the
factor 2 in eq.  (4.6)  and $\sqrt{2}$ in eq.  (4.8)
of  \bggg\  should be omitted  to get the correct
expression.} is obtained from \Wi\ (or \WiA) by interchanging
$\a\leftrightarrow\b$ (or $A\leftrightarrow B$), and setting $p =0$.

There are two simple known solutions of the system \system: the one
found in \refs{\bs,\gpp}, and another one found in \bggg.
Any 7-dimensional manifold $M^7$ with the metric \meet\ has a
particular $U(1)$
isometry acting by the same shift
 on the angular coordinates $\psi$ and $\tilde{\psi}$:
$\psi\to\psi +\nu,\ \tilde{\psi}\to\tilde{\psi} +\nu$.
The field $C(r)$ in \meet\ determines the radius of the
associated circle ( the scale of the $\s_3+\S_3$ direction).
The 11-dimensional manifold of the form $R^{1,3}\times M^7$ can be
reduced  along this $U(1)$
direction to a 10-d background of type IIA
superstring theory. The solutions found in \refs{\bs,\gpp,\bggg}
correspond
after this  reduction to a D6-brane wrapped a three-sphere $S^3$
of deformed conifold
\refs{\amv,\edel}.
Since the type IIA dilaton is given by  $\e^{\p}=C^{3/2}$, the field $C$
determines also the value of the string coupling
(see  \refs{\bggg,\edel} for details).

%%%%%%%%%%%%%%%%%%%%%%%%%%%%%%%%%%%%%%%%%%%%%%%%%%%%%%%%%%%%%%%%%%
\bigskip
{\it Vanishing of correction to warp factor}

\noindent
The  contribution of the
 $R^4$  correction  \JJ\  to the 1-d action corresponding
 to   the ansatz
\metric\  can be obtained using the method
described in section 2. We have checked that the combination
 $I_4(C)$ given by \JJ\
 {\it vanishes}  for {\it any}  solution of the 1-st order
system  \system.
Since  \system\ can be used to express
any derivative of the  metric  and thus its  curvature as  an
algebraic function of $\a,\b,\g,\d$, to check this
 one does not need to know the explicit
form of the solutions of \system.

Since $I_4(C)$ in \JJ\
 depends only on the Weyl tensor, it is easy to see
 that then the correction W$_5^{(1)}$ in \gJ\
 corresponding to the  warp factor $p=\pp^5$
in \metric\  also vanishes.
  This  does not yet imply  that the corresponding
source component $J^5$ in \eqbpsi\  should  also
vanish  since the components  $G^{5a}$ of the metric
are non-trivial.
 Nevertheless,
computing all the corrections W$^{(1)}_a$
% \foot{The explicit form of the
%corrections is complicated, and will not be needed here.}
 and the corresponding  sources $J^a$, we have found
that indeed  $J^5=0$.
As a result, we conclude that
 the $R^4$  correction does not modify the equation
 for $p$, i.e.  we can set $p=0$ so that
the direct product structure of the background
$R^{1,3}\times  M^7$ is preserved at the $R^4$ level.
This is a scheme-dependent statement:
if we used  the action with the Riemann tensor $R_{ijkl}$
instead of the Weyl tensor $C_{ijkl}$ in  $I_4$ in \JJ,
 we would get a nontrivial
correction to the warp factor $p$.
The two schemes are, in general,
 related by a redefinition similar to \chm,
$g_{mn}\to g_{mn} + s_1 \ev  (RRR)_{mn} + s_2 \ev RRR g_{mn} + ...$,
which may rescale  the  11-d metric  by a factor
$1 + k_1 \ev RRR $.

%%%%%%%%%%%%%%%%%%%%%%%%%%%%%%%%%%%%%%%%%%%%%%%%%%%%%%%%%%%%%%%%%%%%%%%%%%%
\subsec{\bf Corrections to BSGPP solution }
%%%%%%%%%%%%%%%%%%%%%%%%%%%%%%%%%%%%%%%%%%%%%%%%%%%%%%%%%%%%%%%%%%%%%%%%

We shall first study  the $R^4$
corrections to the
$G_2$ holonomy metric found in \refs{\bs,\gpp}.
The manifold  has an enhanced
$SU(2)\times SU(2) \times SU(2) \times Z_2$ global
symmetry which is achieved
by setting $\a=\d$, $\b=\g$  in \metrs,
i.e. $D=A$, $C=B$ in \meet.
 Then the system \system\   reduces to
\eqn\systemi{
{dA\ov dr}=  {1\ov 2A}\ ,\ \ \ \ \ \
{dB\ov dr}=  {1\ov 4B}\big(1- {B^2\ov A^2} \big)
\ ,\ \ \ \ \ \ \ \  D=A\ , \ \ \   C=B\ .
}
The general solution
of this leading-order system ($A=A_0, \ B=B_0$) depends on two
parameters $r_0$ and $s$
\eqn\soli{
A_0(r;s,r_0)=\sqrt{r-s}\ ,\ \ \ \ \ B_0(r;s,r_0)=
3^{-1/2}(r-s)^{-1/4}\sqrt{(r-s)^{3/2} - r_0^{3/2}} \ .}
In the special case $r_0=0, \ s=0$
 the metric \meet\ takes the form:
\eqn\meetsp{  \eqalign{
ds_7^2 =&  d \r^2  +  \r^2 \bigg(
  { 1 \ov 12} \big[\big(\s_1-\S_1\big)^2+ \big(\s_2-\S_2\big)^2
 +   \big(\s_3-\S_3\big)^2 \big]
\cr &~~~~~~~~ +\
  { 1 \ov 36}\big[ \big(\s_1+\S_1\big)^2+\big(\s_2+\S_2\big)^2
   +   \big(\s_3+\S_3\big)^2 \big] \bigg) \ , \ \ \ \ \ \ \ \  r\equiv {1 \ov 12} \r^2 \
. }}
This space is a cone which is singular at $\r=0$.
It  follows from the analysis given below that,
like the standard singular conifold,
 this singular  space is not corrected by the $R^4$ term in the
 effective  action.\foot{In contrast to the case of the singular conifold
 the  invariant $E$ \Eul\ is non-zero for this solution.}

The   general non-singular
solution \soli\
can be obtained from the special  solution
with $s=0,\ r_0=1$ by using the translational invariance in $r$
and
the invariance of the  system \systemi\   under the
following scaling transformation:
$
r\to \l^2 r ,\ \ A\to \l A, \ \ B\to \l B$,
which corresponds to the  rescaling of the 7-d metric \meet\
by $\l^2$.
Thus, without loss of
generality,  one can choose $s=0$ and $r_0=1$.
The resulting metric \meet\  (with $1 \leq r < \infty$)
has  regular curvature, and thus
computing corrections in expansion in powers of curvature
is well-defined.

The corrected  analog  \eqbpsi\ of the system \systemi\
is found to be
\eqn\systemic{
{dA\ov dr} =  {1\ov 2A}+\ev J_A\ , \ \ \ \ \ \ \ \ \
{dB\ov dr} =  {1\ov 4B}\big( 1- {B^2\ov A^2}\ \big)+\ev J_B\ ,
}
\eqn\wii{
J_A= 2^{-11}\cdot 3 \ A^{-13}\big( {A}^2 - 3{B}^2 \big)
  \big( 133{A}^4 - 414{A}^2{B}^2 + 301{B}^4 \big)\ ,\ \ \ \ \ \
  J_B= - A^{-1}B J_A \ . }
One is supposed   to solve this system to leading order
in  $\ev$ only (since we ignored higher order in $\ev$ corrections to the
effective action).
 Setting
\eqn\AB{
A=\e^\a =A_0\e^{\ev a}\ ,\ \ \ \  \ \ \ \ \ \  B=\e^\b=B_0\e^{\ev b}\ ,}
where $A_0$ and $B_0$ in \soli\
solve  \systemi,  we find that  $a$ and $b$ should
satisfy
\eqn\systi{
(A_0{d\ov dr}+{dA_0\ov dr} + {1\ov 2A_0}) \ a \ = J_{A_0}\ , \ \ \ \ \ \ \
[B_0{d\ov dr}+{dB_0\ov dr} + {1\ov 4B_0}\big( 1+ {B_0^2\ov A_0^2}
\big)]\ b \  - { B_0\ov 2A_0^2} a
= J_{B_0}\ ,
}
where  $J_{A_0},J_{B_0}$ are given by \wii\ with $A,B \to A_0, B_0$.
The general solution of the inhomogeneous system of  linear equations
\systi\ is given by the  sum of its particular solution and a general
solution  of the homogeneous system obtained by setting the
sources to zero. Since we know the
general two-parameter solution \soli\ of the nonlinear equations \systemi,
we can easily obtain  the
 general solution of the corresponding linearized system
 by differentiating  \soli\ with respect
to the parameters $s$ and $r_0$:
\eqn\sollini{ \eqalign{
a_0(r)&= \c_s {\pa \ln A_0\ov \pa s} +  \c_{r_0} {\pa \ln
A_0\ov \pa r_0} = -{\c_s\ov 2r}\ ,\cr
b_0(r)&= \c_s {\pa \ln B_0\ov \pa s} +  \c_{r_0} {\pa \ln B_0\ov\pa r_0}
= -{\c_s\ov 2r}{r^{3/2}+{1\ov 2}\ov r^{3/2}-1}
-{3\c_{r_0}\ov 4\big( r^{3/2}-1\big)}  \ .}}
Here $\c_s$ and $\c_{r_0}$ are arbitrary constants, and we
have chosen  $s=0$ and
$r_0=1$ here and in what follows (so that   $1 \leq r < \infty$).
Evaluating \wii\ on the solution \soli, one finds
\eqn\wiii{
J_{A_0}= {301 +640 r^{3/2}+256r^3\ov 6144\ r^8}\  , \ \ \ \ \ \ \ \ \
J_{B_0}= -{\sqrt{r^{3/2}-1}\ov
\sqrt{3}\ r^{3/4}}\ J_{A_0} \ .}
Then the general solution of  \systi\ is given by
\eqn\sollsysi{
a(r )=-{\c_s\ov 2r} - {1\ov 84 r^{{9\ov 2}}}- {1\ov 48 r^{6}}
- {301\ov 39936 r^{{15\ov 2}}}\ , }
$$ b(r )= {1\ov r^{{3\ov 2}}-1}\big[ - {\c_s\ov 2}\big( r^{1\ov 2} +{1\ov
2r}\big) -{3\c_{r_0}\ov 4} + {1\ov 63 r^{3}}+ {5\ov 336
r^{{9\ov 2}}} - {823\ov 79872 r^{6}}  -  {2107\ov 299520 r^{{15\ov 2}}}
\ \big]\ .$$
The corrections vanish at large $r$.
We can fix the constant $\c_s$ and $\c_{r_0}$ by imposing the
boundary conditions that
imply that the short-distance ($r\to 1$)
limit  of the metric is also  not changed by the $R^4$ correction,
namely,
\eqn\bcon{
a(1)=0\ ,\ \ \ \ \ \ \  b(1)=0\ , \ \ \ {\rm i.e.} \ \ \ \ \
A(1)=1\ ,\ \ \ \ \ \ \ B(1)=0\ . }
The second condition $b(1)=0$ is a non-trivial one:
 for generic values of
$\c_s$ and $\c_{r_0}$ the function  $b(r)$ has a pole at $r=1$,
so that,  in
general, one could   only require
 regularity of $b(r)$ at this  point.
A simple computation shows that in the present case
 the boundary conditions
\bcon\ can indeed be satisfied,
provided we choose:
\eqn\choos{ \c_s=-  {3753\ov 46592}   \ , \ \ \ \ \ \ \ \ \
\c_{r_0} = {63\ov 640}  \ .
}

\bigskip

%%%%%%%%%%%%%%%%%%%%%%%%%%%%%%%%%%%%%%%%%%%%%%%%%%%%%%%%%%%%%%%%%%%
\subsec{\bf Corrections to BGGG solution}
%%%%%%%%%%%%%%%%%%%%%%%%%%%%%%%%%%%%%%%%%%%%%%%%%%%%%%%%%%%%%%%%%%%
Next, let us
 analyze   corrections to  another  class of  spaces with
$G_2$ holonomy found in \bggg\ (see also \cvet).
 The corresponding metric is given
 by a special solution to the system \system\
depending only on two (out of possible four)
parameters $s,r_0$\foot{The parameter $r_0$  we are using differs from the
one in \bggg\ by a factor of ${3\ov 2}$.}
$$
A_0(r;s,r_0) = \sqrt{
{(r - s - r_0)(r - s + 3r_0)}\ov{8 r_0} }\ ,
\ \ \ \ \
B_0(r;s,r_0) = \sqrt{       {(r - s + r_0)(r - s - 3r_0)\ov 8r_0 }}\ ,
$$
\eqn\solii{
C_0(r;s,r_0) = \sqrt{{2 r_0  (r - s - 3r_0)(r - s + 3r_0)\ov 3
(r - s - r_0)(r - s + r_0)}}\ ,\ \ \ \ \ \ \ \ \ \ \ \
D_0(r;s,r_0) = { r - s \ov \sqrt{6 r_0} }\  .}
As in the BSGPP case,
the presence of the  two free parameters
$s$ and $r_0$ simply
reflects the invariance of \system\ under a shift
 and a rescaling of the
coordinate $r$. In what follows we shall
set $s=0$ and $r_0=1$, so that the range of $r$ will
be $ 3 \leq r < \infty$.

To find the corrections  we need to know
the general solution of  the homogeneous system of linear equations
describing small perturbations of $A,\ B,\ C$ and $D$ around the
 solution
\solii, as well as
 a particular solution of the inhomogeneous system with
sources corresponding to  $R^4$ correction
expanded near the solution \solii.
Unfortunately, in the present case
it does not seem possible
 to find the exact general  solutions of these two systems.
  That  makes
the computation of the corrections much more complicated.
Instead of finding the  solutions  numerically,
 we will determine  them
in  the vicinity of $r=3$ by  expanding
 in powers of $r-3$, and also at large $r$ by expanding
 in powers
of $1/r$. We will then  sew the two expansions.
 This will allows us to
get all essential information  for determining
the corrections, though the explicit   analytic
form of the corrected solution  in this case will not be available.

\bigskip
%%%%%%%%%%%%%%%%%%%%%%%%%%%%%%%%%%%%%%%%%%%%%%%%%%%%%%%%%%%%%%%%%%
{\it Linearized  perturbations near the leading solution }
%%%%%%%%%%%%%%%%%%%%%%%%%%%%%%%%%%%%%%%%%%%%%%%%%%%%%%%%%%%%%%%%%%

\noindent
Let us start with the system of linear equations describing
small perturbations of $A,\ B,\ C$ and $D$ around
the solution \solii.\foot{The analysis  below extends and corrects the previous discussion in \bggg.}
Representing the fields as in \AB,
\eqn\ABCD{
A=\e^\a =A_0\e^{\ev a}\ ,\ \ \ \   B=\e^\b =B_0\e^{\ev b}   \ ,\ \ \ \
C=\e^\g =C_0\e^{\ev c}\ ,\ \ \ \   D=\e^\d =D_0\e^{\ev d}
\ ,}
we find the  system of  equations for the perturbations
near \solii\ with $s=0, \ r=1$ (cf. \systi)
$$
{d a\ov dr} =-{5r^2-12r+3\ov 2r(r-1)(r-3)}a +
{b\ov 2r}  -{c\ov r+3} + {d\ov r-3} \ ,\ $$
\eqn\sysliii{ {d b  \ov dr}  =
{a\ov 2r} -{5r^2+12r+3\ov 2r(r+1)(r+3)}b -{c\ov r-3} + {d\ov r+3}\ ,  }
$$
{dc\ov dr} ={4a\ov\ (r-1)(r+3)}-{4b\ov\ (r-1)(r+3)}  ,\
\ \ \
 {d \ d \ov dr}=
{2a\ov r-3}  +{2b\ov r+3} - {c\ov r}-{5r^2-9\ov r(r^2-9)}d \ . $$
%%%%%%%%%%%%%%%%%%%%%%%%%%%%%%%%%%%%%%%%%%%%%%%%%%%%%%%%%%%%%%%%%%%%
%$$
%({d \ov dr} +{1\ov r-3} + {1\ov r-1} + {1\ov 2r}) a -
%{b\ov 2r}  +{c\ov r+3} - {d\ov r-3} =0\ ,\ $$ \eqn\sysliii{
%({d   \ov dr} +{1\ov r+3} + {1\ov r+1} + {1\ov 2r})b -
%{a\ov 2r}  +{c\ov r-3} - {d\ov r+3} =0\ ,  }
%$$
%{d\ov dr}c +{a\ov r+3} - {a\ov r-1} +{b\ov r-3} - {b\ov r+1} =0\ ,\
%\ \ \
%( {d  \ov dr} +{2\ov r+3} + {2\ov r-3} + {1\ov r})d  -
%{2a\ov r-3}  -{2b\ov r+3} + {c\ov r} =0\ . $$
%%%%%%%%%%%%%%%%%%%%%%%%%%%%%%%%%%%%%%%%%%%%%%%%%%%%%%%%%%%%%%%%%%%%%%
Differentiating the solution \solii\ with respect to the
parameters $s$ and $r_0$ (and then
setting them  equal
to $s=0, \ r_0=1$) as in \sollini, we
obtain  the following special two-parameter solution to \sysliii\foot{Here
$a_0(r)\equiv -\c_s {\pa \ln A_0\ov \pa s} + 2 \c_{r_0} {\pa \ln
A_0\ov \pa r_0}$, etc.}
 $$
a_0(r) = \c_s\frac{r+1}{(r -1)(r+3) }
 -\c_{r_0}\frac{r^2+3}{ (r -1)(r+3) }, \ \ \
b_0(r)=   \c_s\frac{r-1}{(r +1)(r-3) }
 -\c_{r_0}\frac{r^2+3}{ (r +1)(r-3) }  , $$
 \eqn\sollinii{
c_0(r)=  \c_s\frac{8r}{(r^2 -1)(r^2-9) }
 +\c_{r_0}\frac{r^4-26r^2+9}{ (r^2 -1)(r^2-9) } \ ,  \ \ \ \ \ \
d_0(r)= {\c_s\ov r}  - \c_{r_0} \ .  }
The system of four linear equations \sysliii\
must have  four independent
solutions. Since  we do not know how to find
the remaining two solutions
 in a closed form, we shall
  analyze them in the vicinity of $r=3$
and at large $r$, and then sew the solutions obtained.
Let us write the  \sysliii\ in the form
\eqn\sysMform{
{d\p^s\ov dr}=M^s_{\ q}(r) \p^q\ ,  \ \ \ \ \ \ \ \ \
\phi^q= (a,b,c,d)  \ .}
The  matrix $M(r)$ has the following expansion near $r=3$
\eqn\eexx{
M(r) = {\M\ov r-3} + O(1)\ ,\ \ \ \ \ \ \  \ \
\M = \pmatrix{
-1 & 0 & 0 & 1  \cr
0 & 0 & -1 & 0  \cr
0 & -1 & 0 & 0  \cr
2 & 0 & 0 & -2 } \ .
}
$\M$ has the following
  eigenvalues $\l_q$ and the
corresponding eigenvectors $v_q$
\eqn\eigenvv{ \eqalign{
&\l_1=-3\ ,\ \ \ v_1=(-1,0,0,2)\ ; \ \ \ \ \ \ \ \ \
 \l_2=-1\ , \ \ \ v_2= (0, 1, 1, 0)\
; \cr
&\l_3=0\ , \ \ \ v_3= (1, 0, 0, 1)
\ ;\ \ \ \ \ \ \ \ \ \  \ \ \ \ \l_4=1\ , \ \ \ v_4=  (0, -1, 1, 0) \ .} }
Near $r=3$ the solution corresponding to an eigenvalue $\l$ behaves
 as $(r-3)^\l [1+O(r-3)]$.
The solutions corresponding to $\l_2=-1$ and $\l_3=0$
 can be  obtained from the
two-parameter solution \sollinii\ by choosing
$\c_s=2\ ,\ \c_{r_0}=2/3$ and $\c_s=6\ ,\ \c_{r_0}=1$,
respectively. Thus we know their exact form (away from $r=3$)\foot{We  will denote
 the solution corresponding to the
eigen-value $\l_k$ as $(a_k,b_k,c_k,d_k)$.}
$$
a_2 =
-{2r(r-3)\ov 3(r-1)(r+3)}\ ,\ \ \ \ \ \ \ \ \
b_2 =-{2(r^2-3r+6)\ov 3(r+1)(r-3)}\ , $$  \eqn\sols{
c_2 ={2(r^4-26r^2+24r+9)\ov 3(r^2-1)(r^2-9)}\ ,\ \ \ \ \ \ \
d_2 =-{2(r-3)\ov 3r }\ ,}
and
\eqn\solss{
a_3 =
-{r^2-6r-3\ov (r-1)(r+3)}\ , \ \ \
b_3 =-{r-3\ov r+1}\ ,\ \ \ \
c_3 ={(r-3)(r^2+6r+1)\ov (r^2-1)(r+3)}\ , \ \ \
d_3 ={6-r\ov r }\ .}
Note that, contrary to the claim  in \bggg, the
perturbation \solss\ does not describe a new deformation of the
two-parameter solution \solii\ but corresponds simply
 to a change of the
parameters $s$ and $r_0$.

The perturbation corresponding to
 $\l_1=-3$ is singular at $\tau \equiv r-3=0$ (i.e. it does not represent
a smooth
deformation of the solution \solii); it is found to be
 $$
a_1 = -{1\ov \tau^{3}} + \frac{2}{3\tau^2} - \frac{7}{36\tau} - \frac{2\tau}{27}
+ \frac{661\tau^2}{12960}  + O(\tau^3)+\frac{\big( 4 + \tau \big) \ln \tau}
   {3\big( 2 + \tau \big) \big( 6 + \tau \big) } \ ,\  \ \ \ \ \tau\equiv r-3\ , $$
   $$
b_1 =  \frac{17\tau}{216} -\frac{17}{72}  -
\frac{37\tau^2}{1296} + O(\tau^3) + \frac{\big( 2 + \tau \big)
\ln \tau}    {3\tau\big( 4 + \tau \big) } \ ,  $$  $$
c_1 =   \frac{1}{6\tau^2} - \frac{5}{18\tau}+ \frac{5}{108}-
 \frac{5\tau^2}{288} + O(\tau^3)+ \frac{8\big( 3 + \tau \big) \ln \tau}
   {3\tau\big( 2 + \tau \big) \big( 4 + \tau \big) \big( 6 + \tau \big) }
   \ ,  $$ \eqn\solmthr{
d_1 = \frac{2}{\tau^3} - \frac{4}{3\tau^2} + \frac{7}{9\tau}-\frac{5}{12}
 +  \frac{55\tau}{648} - \frac{13\tau^2}{1215} + O(\tau^3) +
  \frac{\ln \tau}{3\big( 3 + \tau \big) } \ . }
The  $\ln \tau$ terms  in \solmthr\ (which are exact)
are multiplied by the functions of the explicitly known
two-parameter solution \sollinii\ with $\c_s=1/3, \
\c_{r_0}=0$.

The fourth linearized solution  corresponding to
$\l_4=1$ vanishes at $r= 3$  and (as we shall see below)
 goes
to a constant at $r=\infty$. Since the  value of the
function $C$ in \meet\ may be interpreted as in \bggg\  as
determining  the radius of the M-theory circle,
{\it  this}   perturbation
describes a {\it nontrivial deformation }
of the manifold  corresponding to a
finite change in this  radius at infinity \amv\ (cf. \bggg).
 Near  $\tau =r-3=0$ it  is given by  a series in powers of $\tau$
with the  radius of convergence $|\tau | < 2$. To match the solution to  a
solution found at large $r$ we computed it up to the order $\tau^{20}$.
The first few terms of the series are
 $$
a_4 =-\frac{\tau^2}{5} + \frac{2\tau^3}{15} - \frac{131\tau^4}{1890}+ O(\tau^5)\
,\ \ \ \ \ \ \
b_4 = -\tau + \frac{17\tau^2}{36} - \frac{29\tau^3}{144} +
\frac{1141\tau^4}{14400} + O(\tau^5)\ , $$ \eqn\solpone{
c_4 = \tau - \frac{13\tau^2}{36} + \frac{227\tau^3}{2160} -
\frac{103\tau^4}{4800} + O(\tau^5)\ ,\ \ \ \ \ \ \
d_4 = -\frac{4\tau^2}{15} + \frac{8\tau^3}{45} - \frac{244\tau^4}{2835}+
O(\tau^5)\ .}
Next, let us solve  the system  \sysliii\ at
large $r$, where
$M(r)$ in \sysMform\  has the form
\eqn\huh{
M(r) = {\M_{\infty}\ov r} + O({1\ov r^2})\ ,\ \ \ \ \ \ \ \
\M_{\infty} = \pmatrix{
-{5\ov 2} & {1\ov 2} & -1 & 1  \cr
{1\ov 2} & -{5\ov 2} & -1 & 1  \cr
0 & 0 & 0 & 0  \cr
2 & 2 & -1 & -5 } \ .
}
The eigenvalues  and  eigenvectors  of $\M_\infty$ are
\eqn\eigenvv{ \eqalign{
&\l_1=-6\ ,\ \ \ v_1=(1,1,0,-4)\ ;\ \ \ \ \ \ \ \ \  \l_2=-3\ , \ \ \
v_2= (1, -1, 0, 0)\ ; \cr
&\l_3=-1\ , \ \ \ v_3= (1, 1, 0, 1)
\ ; \ \ \ \ \ \ \ \ \ \ \  \l_4=0\ , \ \ \ \  v_4=  (-1, -1, 1, -1) \ .} }
Here the  solution associated  to an eigenvalue $\l$ goes
 as $r^\l [1+O({1\ov r})]$.
The solutions corresponding to
$\l_3=-1$ and $\l_4=0$ are obtained from the
two-parameter solution \sollinii\ by imposing the relations
$\c_s=1\ ,\ \c_{r_0}=0$ and $\c_s=0\ ,\ \c_{r_0}=1$,
respectively. The solution corresponding to  $\l_1=-6$ is
given by a power series in  $1/r$.
The series converges very slowly for $r > 3$, and to match the solutions
with the ones found  near  $r=3$ we computed the terms  up to
the order $1/r^{30}$. The first few terms are
$$
a_{1\infty} =   \frac{1}{r^6}+ \frac{4}{r^7}+ \frac{205}{7 r^8}
 + \frac{754}{7 r^9} +O({1\ov r^{10}})\ ,\ \ \ \ \
b_{1\infty} =  \frac{1}{r^6} - \frac{4}{r^7} + \frac{205}{7 r^8}
- \frac{754}{7 r^9}+O({1\ov r^{10}})\ , $$
\eqn\solmsix{
c_{1\infty} =  - \frac{2}{r^8} +O({1\ov r^{10}}) \
,\ \ \ \ \ \ \
d_{1\infty} = - \frac{4}{r^6} - \frac{810}{7 r^8} +O({1\ov r^{10}}) \ .   }
The $1/r$ expansion for the solution
for $\l_1=-3$ breaks down at the order $1/r^6$
where a $ {1\ov r^6}\ln r $ term appears.
The  coefficients of the $\ln r$ terms are
 proportional   to the solution \solmsix\ (cf. \solmthr), i.e.
  $$
a_{2\infty} = \frac{1}{r^3}- \frac{2}{r^4}- \frac{5}{r^5}
 +O({1\ov r^{7}})+ {324\ov 5}a_{1\infty} \ln r \ ,\ \ \ \ \ \
b_{2\infty} = -\frac{1}{r^3} -\frac{2}{r^4}+ \frac{5}{r^5}
 + O({1\ov r^{7}})+ {324\ov 5}b_{1\infty } \ln r \ ,  $$
\eqn\solmthre{ c_{2\infty} = -\frac{2}{r^4} -\frac{8}{r^6}
+ O({1\ov r^{8}})+ {324\ov 5}c_{1\infty} \ln r \ ,\ \ \ \ \
d_{2\infty} = \frac{6}{r^4}-\frac{136}{5r^6}
+ O({1\ov r^{8}})+ {324\ov 5}d_{1\infty } \ln r \ .}
Now we are ready to determine the large $r$ behavior of the nontrivial
solution \solpone\ vanishing at $r=3$. At large $r$ it is
represented  by a
superposition of the independent solutions \sollinii, \solmsix\ and
\solmthre, i.e.
\eqn\match{
\p^i_4= \p_0^i(r) + \c_1\p_{1\infty}^i(r) + \c_2\p_{2\infty}^i(r)\ ,}
where $\ \p_0^i(r),\ \p_{1\infty}^i(r)$ and  $\p_{2\infty}^i(r)$  are
given by  \sollinii, \solmsix\ and \solmthre, respectively.
Computing the l.h.s and the r.h.s. of \match\ at different values of
$r=\tau +3$, we find the constants $\c_{r_0}$, $\c_s$, $\c_1$ and $\c_2$
\eqn\consts{
\c_{r_0}= -2.00\ \pm\ 0.02\ , \ \ \ \c_s=-9.31\ \pm\ 0.09\ ,
\ \ \ \c_1=-7050\ \pm\ 70\ , \ \ \ \c_2=54.3\ \pm\ 0.5\ .}
These values  were obtained by matching the
solutions at $r = 17/4$.

In the process of the analysis of small perturbations near the special
solution \solii\ we have thus shown that
the system \system\ has a  {\it three}-parameter family of {\it regular}
 solutions, with the third non-trivial parameter
corresponding to the perturbation $\p^i_4$.
The existence of a 3-parameter family
of solutions of the 1-st order system  \system\
 was also demonstrated in \cvet\ by using
numerical analysis of \system.
%\foot{Note that
%since the radius $R_\infty (\ev )$ of the  $U(1)$
%circle at $r=\infty$ is equal to $C(\infty )$, where $C(r)$ is given by
%\ABCD, and is determined by the value of the constant $\c_{r_0}$, we thus
%find that $ R_\infty (\ev )/R_\infty (0)= 1-2\ev $.}
%%

%%%%%%%%%%%%%%%%%%%%%%%%%%%%%%%%%%%%%%%%%%%%%%%%%%%%%%%%%%%%%%%%%%%%%

\bigskip
{\it Solution of the inhomogeneous system}
%%%%%%%%%%%%%%%%%%%%%%%%%%%%%%%%%%%%%%%%%%%%%%%%%%%%%%%%%%%%%%%%%%%%%

\noindent
Now we are ready to study the  $R^4$ corrections to the solution \solii.
Computing the source terms $J^s$ in the corrected
analog  \eqbpsi\ of \sysliii\  from   \JJ\ expanded \gJ\ near  the
leading-order
solution \solii, we find the following inhomogeneous
system of linear equations that replaces \sysMform\
\eqn\sysinh{
{d\p^s\ov dr}=M^s_{\ q}(r) \p^q + J^s(r)\ , }
where the matrix $M(r)$ is the
same as in \sysMform, and the sources $J^s$ are given by\foot{We first found
$J^s$ for an arbitrary solution of \system\ as  functions of
$\p^s$ by using the method described in section 2, and
 then  computed them
on the special solution  \solii\ with $s=0$ and $r_0=1$.}
$$
J^1= -\frac{1024\big( -78 - 405r - 286r^2 -
549r^3 - 58r^4 + 57r^5 - 26r^6 +
      r^7 \big) }{27{\big(  r^2 -1 \big) }^8} \ ,  $$
      \eqn\sources{
      J^2= -\frac{1024\big( 78 - 405r + 286r^2 -
549r^3 + 58r^4 + 57r^5 + 26r^6 +
      r^7 \big) }{27{\big(  r^2 -1 \big) }^8} \ ,  }
      $$
J^3= -\frac{2048r\big( 577 + 1089r^2 - 101r^4 + 3r^6 \big) }
  {27{\big(  r^2 -1 \big) }^8}   \ ,\ \ \ \
J^4= \frac{2048r\big( -233 - 9r^2 + 13r^4 + 5r^6 \big) }
  {27{\big(  r^2 -1 \big)  }^8}\ .  $$
To solve the resulting system
we  follow the same strategy that
we used to analyze the small perturbations near the leading solution:
find approximate solutions of  \sysinh\ near
$r=3$ and at large $r$, and then sew the two expansions.

To solve  \sysinh\ near  $r=3$ we impose the
conditions
\eqn\bcs{
\p^s (\tau )|_{\tau =0}=0\ , \ \ \ \ \ \ \  \tau \equiv r-3 \ .   }
As follows from the analysis of small
perturbations, this condition fixes the  solution modulo  the solution
\solpone\ of the system \sysMform\ which also vanishes at $\tau =0$.
We computed the expansion in $\tau $
up to the order $\tau^{20}$. The series has a radius of convergence
$|\tau | < 1$, and its first few terms are given by
 $$
a(\tau)= \frac{137 \tau}{2304} - \frac{617 \tau^2}{3840}  + O(\tau^3) +
\c_4 a_4(\tau)\ ,\ \ \ \ \ \ \
b(\tau)= -\frac{137\tau}{2304} + \frac{5933\tau^2}{82944} + O(\tau^3) +
\c_4 b_4(\tau)\ ,  $$ \eqn\solinhu{
c(\tau)= \frac{11213\tau^2}{82944} + O(\tau^3) +
\c_4 c_4(\tau)\ ,\ \ \ \ \ \ \ \
d(\tau)= \frac{137\tau}{2304} - \frac{5189\tau^2}{34560}  + O(\tau^3) +
\c_4 d_4(\tau)\ .}
Here $\c_4$ is an arbitrary constant which is not fixed by the boundary
conditions, and $a_4,b_4,c_4,d_4$ is the solution \solpone\
of the  homogeneous system. We will later fix the constant $\c_4$ by
requiring that the corrected solution also vanishes at $r=\infty$.
This is a natural requirement since corrections to the BGGG background
vanish at large $r$.

Now let us study the large $r$ region.
Since the sources $J^q$ go to zero  at large $r$ as
$1/r^9$,  there is a solution starting with $1/r^8$. The series converges
very slowly for $r >3$,  and we computed it up to the order $1/r^{40}$.
The leading terms are
 $$
a_\infty (r) =  \frac{7424}{189r^8}- \frac{12800}{189r^9}
+ O({1\ov r^{10}})\ ,\ \ \ \ \ \ \
b_\infty (r) = \frac{7424}{189r^8}+ \frac{12800}{189r^9}
+ O({1\ov r^{10}})\ ,\ \ \ \ \  $$ \eqn\solinhr{
c_\infty (r) =   \frac{256}{9 r^8}
+ O({1\ov r^{10}})\ ,\ \ \ \ \ \
d_\infty (r) = - \frac{32000}{189 r^8}
+ O({1\ov r^{10}})\ .}
Now it is straightforward to find the large $r$ behavior of the solution
\solinhu\ vanishing at $r=3$. At large $r$ it is given by a sum of the
solution \solinhr\ and a solution of the homogeneous system \sysMform,
i.e.
\eqn\matchin{
\p^q = \p^q_\infty (r)+\p_0^q(r) + \c_1\p_{1\infty}^q(r) +
\c_2\p_{2\infty}^q(r)\ ,}
where $ \p^q_\infty(r),\ \p_0^q(r),\ \p_{1\infty}^q(r)$
and  $\p_{2\infty}^q(r)$ are given by  \solinhr, \sollinii, \solmsix\ and
\solmthre, respectively. Computing the l.h.s and the r.h.s. of \matchin\ at
different values of $r=\tau +3$, we find that the constants $\c_{r_0}$,
$\c_s$ in \sollinii\ and
 $\c_1$ and $\c_2$ are expressed in terms of the constant $\c_4$ as
follows \eqn\constsin{ \eqalign{
\c_{r_0}=0.10\ \pm\ 0.01\  -\ (2.0\ \pm\ 0.2)\c_4\ &, \ \ \
\c_s=0.51\ \pm\ 0.05\ -\ (9\ \pm\ 1)\c_4 \ ,\cr
\c_1=320\ \pm\ 30\ -\ (7000\ \pm\ 700)\c_4\ &, \ \ \
\c_2=-2.5\ \pm\ 0.3\ +\ (54\ \pm\ 5)\c_4\
.}}
These values  were obtained by sewing
the solutions at $\tau  =99/100$. Comparing these  with \consts, we see
that they match. We also see that if one chooses $\c_4 \approx 0.05$ then
the coefficient $\c_{r_0}$ vanishes, and, therefore, the
corrections to the metric vanish not only at $r=3$ but also at
 $r=\infty$.
 The resulting corrected solution then smoothly interpolates
 between the same short-distance and large-distance
 asymptotic values.

%Although for generic values of $\c_4$ the
%coefficient $\c_{r_0}$ does not vanish, and, therefore, the $R^4$
%corrections change the radius of the $M$-theory  circle at infinity,
%the correction to the radius
%vanishes at large $r$ if one chooses $\c_4 \approx 0.05$.

%%%%%%%%%%%%%%%%%%%%%%%%%%%%%%%%%%%%%%%%%%%%%%%%%%%%
%\bigskip
\subsec{\bf Corrections to $G_2$ spaces as solutions of  10-d
 superstring theory}

It is straightforward to consider
the leading $R^4$ corrections to the 10-d  backgrounds of the
form  $R^{1,2}\times M^7$ in type II superstring theory,
in the same way as this  was done for the conifolds in section 3.
One can check that with  the scheme choice  corresponding to \acor\
 \  (i) the direct
product structure of the manifold $R^{1,2}\times M^7$ is
again preserved, i.e. the warp-factor is zero; \ (ii)
the metric of  $M^7$ space   gets the same    corrections
as in 11 dimensions; (iii) as in the conifold case
(see section 3.2) the dilaton is again  shifted by
${1\ov 3}\ev E$, where $E$ is the
cubic curvature  invariant in \Eul\
(so that there is a scheme where the dilaton is unchanged).
The explicit expressions for $E$
for the two spaces \soli\ and \solii\  are ($s=0,\ r_0=1$)
$$
E_{BSGPP}= -\frac{ 553 + 915\,r^{\frac{3}{2}} + 480\,r^3 +
320\,r^{\frac{9}{2}} }{3072\,r^{\frac{15}{2}}}  ,
\ \ \ \
E_{BGGG}= \frac{512\,\left( 99 + 207\,r^2 + 9\,r^4 + 5\,r^6
\right)}
{9\,{\left( r^2-1 \right) }^7} .
$$
One may ask about the connection between the 10-d and 11-d results.
Earlier in this section we have shown
that the direct product
structure of  the  space $R^{1,3}\times M^7$
 is preserved by the $R^4$  corrections to the effective action
 in 11 dimensions.
 If we reduce the 11-d background
 to 10 dimensions
  along  one of the  ``free''  spatial directions of
 $R^{1,3}$  we get a 10-d type IIA string  background
 of  the form $R^{1,2}\times M^7$ with  {\it constant} dilaton.
Thus, the $R^4$  corrections \JJ\ in 11
dimensions reduced to 10 dimensions should
give the (one-loop)  $R^4$ corrections in
10 dimensions in the scheme where
 the dilaton is not modified, i.e.
where there is no  $E\nn^2\p$ term in  \Lonen.

\bigskip
\noindent
{\bf Acknowledgements}
%%%%%%%%%%%%%%%%%%%%%%%%%%%%%%%

\noindent
We are grateful to C. Nunez for a collaboration at  an initial
 stage and  many  discussions.
We thank  G. Papadopoulos and S. Shatashvili 
 for useful  correspondence.
 We also acknowledge Yu. Obukhov for help 
 with GRG program. 
The work of S.~F. and A.~A.~T.  was supported by
the DOE grant
DE-FG02-91ER40690.
The work of A.~A.~T.  was also supported in part by
the  PPARC SPG 00613,
 INTAS  99-1590 and  CRDF RPI-2108 grants.

%%%%%%%%%%%%%%%%%%%%%%%%%%%%%%%%%%%%
\appendix{A}
{Identity for $E_{ij}$ and 11 $\to$ 10 dimensional
reduction of  $R^4$ terms }
%%%%%%%%%%%%%%%%%%%%%%%%
Here we shall first
prove the identity \ijEij\  for the $E_{ij}$
tensor defined in \Eij. We shall   then comment on
the role of this identity in checking the consistency
of the relation by dimensional reduction between the $R^4$ terms
in the M-theory action
 and the  superstring effective action, which is
implied by local supersymmetry.

Let us first recall the origin of $E_{ij}$:
it  appears   from the $R^4$ invariant \jnu\  upon performing
a conformal
variation of the metric: $\delta g_{ij} = \psi g_{ij} $,
\  $ \delta I_4(R) = 4 E^{ij} \nabla_i \nabla_j \psi$.
Since the structure of $I_4(R) $  is strongly constrained
by supersymmetry, the same should be true for  $E_{ij}$.
We suspect that there may be a way to
understand  the existence of the identity \ijEij\
from the fact that the structure of $E_{ij}$ is constrained
by the  local supersymmetry.

 First, it is easy to check that if \ijEij\ holds for
 any Einstein manifold, then the same identity should hold also
 with the Weyl tensors in \Eij\ and \Eul\ replaced  by
the Riemann ones.  Then omitting terms that vanish for $R_{ij} = k g_{ij}$
we obtain
%\foot{We do not distinguish upper and lower indices
%here.}
 \eqn\jEij{
\nn_i\nn_j E_{ij}-{1\ov 6}\nn_i\nn_i E =I_1+I_2+I_3+I_4\ ,}
where
\eqn\Ii{\eqalign{
I_1&=-\nn_i\nn_j R_{imkl}R_{jpkq}R_{lpmq}
+{1\ov 4}\nn_i\nn_j R_{imkl}R_{jmpq}R_{klpq}\cr
&+{1\ov 2}\nn_i\nn_i R_{jmkl}R_{mpql}R_{pjkq}
+{1\ov 4}\nn_i\nn_i R_{jmkl}R_{jmpq}R_{pqlk}\ ,}}
\eqn\Iii{\eqalign{
I_2&=
-R_{imkl}R_{jpkq}\nn_i\nn_jR_{lpmq}
+{1\ov 4}R_{imkl}R_{jmpq}\nn_i\nn_j R_{klpq}\cr
&+{1\ov 2}R_{kijl}\nn_i\nn_j R_{kmpq}R_{lmpq}
+{1\ov 2}R_{lijk}\nn_i\nn_j R_{kmpq}R_{lmpq}\ ,}}
\eqn\Iiii{
I_3={1\ov 4}\nn_j R_{imkl}\nn_i\left( R_{jmpq}R_{klpq}\right)
-{1\ov 4}\nn_i R_{jmkl}\nn_i\left( R_{jmpq}R_{klpq}\right)\ ,}
\eqn\Iiiii{\eqalign{
I_4&=
-\nn^jR_{imkl}\nn^i \left( R_{jpkq}R_{lpmq}\right)
-R_{imkl}\nn^i R_{jpkq}\nn^j R_{lpmq}  \cr
&+{1\ov 4}R_{imkl}\nn_i R_{jmpq}\nn_j R_{klpq}
+{1\ov 2}\left( R_{kijl}+ R_{lijk}\right)\nn_j R_{kmpq}\nn_i R_{lmpq}\cr
&+ {1\ov 2}\nn_i R_{jmkl}\nn_i R_{mpql}R_{pjkq}
+ {1\ov 2}\nn_i R_{jmkl}R_{mpql}\nn_iR_{pjkq}\ .}}
We are going to show that $I_1=0, \ I_2=0, \ I_3=-I_4$, thus
demonstrating that the r.h.s. of \jEij\ is zero.

Using  the Bianchi identity we find
$
{1\ov 2}\nn_i\nn_i R_{jmkl}R_{mpql}R_{pjkq}=
\nn_i\nn_j R_{imkl}R_{jpkq}R_{lqmp},$

\noindent
$
{1\ov 4}\nn_i\nn_i R_{jmkl}R_{jmpq}R_{pqlk}=
{1\ov 2}\nn_i\nn_j R_{imkl}R_{jmpq}R_{lkpq} \ .
$
Using then  cyclic identity, we get
$$
I_1= \nn_i\nn_j R_{imkl}R_{jpkq}R_{lmqp}
-{1\ov 4}\nn_i\nn_j R_{imkl}R_{jmpq}R_{klpq}\ .
$$
The cyclic identity gives also
$
\nn_i\nn_j R_{imkl}R_{jpkq}R_{lmqp}=
{1\ov 2}\nn_i\nn_j R_{ikml}R_{jmpq}R_{klpq}\ ,
$
so that finally  $I_1=0$.
To show that $I_2=0$ we need the following identities
$$
R_{kijl}\nn_i\nn_j R_{kmpq}R_{lmpq}=
R_{kiml}\nn_i\nn_j R_{kmpq}R_{ljpq}
- {1\ov 2}R_{ijkl}[\nn_i,\nn_j] R_{lpmq}R_{mqpk}\ ,$$
$$
R_{lijk}\nn_i\nn_j R_{kmpq}R_{lmpq}=
 {1\ov 2}R_{limk}\nn_i\nn_j R_{kmpq}R_{ljpq}
\ , $$ $$ \ \ \ \
\nn_i\nn_j R_{lpmq}R_{ipkl}R_{jkmq}=2 \nn_i\nn_j
R_{lpmq}R_{imkl}R_{jkpq}\ ,$$
and
$
\nn_i\nn_j R_{mqlp}R_{imkl}R_{jqpk}={1\ov 4}R_{ijkl}[\nn_i,\nn_j]
R_{lpmq}R_{mqpk}\ .$
It is easy to see that
$$
I_3= -{1\ov 4}\nn_j R_{imkl}\nn_i\left( R_{jmpq}R_{klpq}\right)\ .
$$
Since \
$
{1\ov 2}\left( R_{kijl}+ R_{lijk}\right)\nn_i R_{kmpq}\nn_j R_{lmpq}=
 \nn_i R_{jqkm} \nn_j R_{lmpq}\left( R_{kipl}+ R_{lipk}\right) , $
we get
$$
I_3+I_4= \nn_i R_{jpkq}\big( \nn_j R_{imkl} R_{lpmq}
-{1\ov 4}\nn_j R_{ipml} R_{kqml}+\nn_j R_{qmpl} R_{ilmk}
-\nn_i R_{jmkl} R_{lqpm}\big).$$
This is simplified using
$
-{1\ov 4}\nn_i R_{jpkq}\nn_j R_{ipml} R_{kqml}=
{1\ov 2}\nn_i R_{jpkq}\nn_j R_{pqlm} R_{ikml}$ and

\noindent
$
\nn_i R_{jpkq}\nn_i R_{jmkl} R_{lqpm}=
\nn_i R_{jpkq}\big(\nn_j R_{imkl} R_{lqpm}
- \nn_j R_{pmql} R_{imkl}\big).$
Then finally
 $$
I_3+I_4= \nn_i R_{jpkq}\nn_j R_{pmkl} R_{ikml}
+ {1\ov 2}\nn_i R_{jpkq}\nn_j R_{pqlm} R_{ikml}=0\ .$$
This completes the proof of
\eqn\kkk{
\nn^i\nn^j E_{ij}={1\ov 6}\nn^2 E\ . }
As is well known, the 11-d and 10-d supergravities
are related by dimensional reduction along an isometric direction.
Since the  $R^4$ invariants \acor\ in 10 and  \acorG\ in 11 dimensions
should be consistent  with the respective
supersymmetries, one expects them to be  also  related  by dimensional
reduction. While this is obviously true for the purely  metric-dependent
terms,  let us  compare the  simplest dilaton-dependent
terms $R^3\nn^2\p$  as they appear upon dimensional reduction
from  \acorG\  with the similar terms present in the string
effective action  \acor.\foot{The 11-d $R^4$ term is related
to string one-loop $R^4$ term whose form  is different
from the tree-level invariant  \acor\  by the $\ee_{10}\ee_{10}RRRR$
term.
However, this difference is irrelevant in the present case as
 the term
$\ee_{10}\ee_{10}RRR\nabla^2\phi$ vanishes due to  Bianchi identity,
and thus $C^3\nabla^2\phi$ term
following from \acor\ coincides with the
corresponding one-loop term up to contributions that
 vanishing on the leading-order equations of motion.}
As usual, we shall assume that
the 11-d metric can be written as
$$
ds_{11}^2 = \e^{-\p/6}ds_{10E}^2 + \e^{4\p/3}(dx_{11} + C_m dx^m )^2\ ,$$
where $\p$ and $C_m$ are the dilaton and the RR  vector field,
respectively, and  $ds_{10E}^2$ is the 10-d metric in the
Einstein frame.
Since we want to consider the linear dilaton term
 $C^3\nabla^2\p$ that follows from
  the $C^4$ term in 11 dimensions, we
can  consider only the components of the Weyl tensor
 that have  10-dimensional
indices. Modulo  field redefinition ambiguity we can replace the
Weyl tensor $C_{ijkl}$ by the Riemann one $R_{ijkl}$.
Using the general relation between curvatures of the two
conformally-equivalent  spaces ($ \tilde{g}_{ij}=\e^{2\pp}g_{ij}$)
$$
\tilde{R}^i{}_{jkl}=  R^i{}_{jkl}-\d^i_k ( \nabla_j\nabla_l\pp -
\nabla_j\pp\nabla_l\pp )+ \d^i_l ( \nabla_j\nabla_k\pp -
\nabla_j\pp\nabla_k\pp )- g_{jl}( \nabla^i\nabla_k\pp -
\nabla^i\pp\nabla_k\pp )$$
$$+\ g_{jk}( \nabla^i\nabla_l\pp -
\nabla^i\pp\nabla_l\pp ) -\nabla^m\pp\nabla_m\pp (\d^i_k g_{jl}-
\d^i_l g_{jk} )\ , $$
we derive the following relation between 11-d and
10-d Riemann tensors
\eqn\relll{
R^{(11)}{}^m{}_{nrs}= R^{(10)}{}^m{}_{nrs}
+{1\ov 12}\left( \d^m_r\nabla_n\nabla_s \p
-\d^m_s\nabla_n\nabla_r\p
-\d^n_r\nabla_m\nabla_s\p +\d^n_s\nabla_m\nabla_r\p \right)\ ,}
where we have  omitted  terms quadratic in $\p$ and terms
proportional to the dilaton
equation of motion (i.e. $\nabla^m \nabla_m \p$).
 Comparing \relll\  with \Cbar, we conclude  that the
coefficients in front of the tensor $(\nabla^2\p)^m{}_{nrs}$ differ by
the factor $-{1\ov 3}$. Therefore, if we
 start from the 11-dimensional $R^4$ term and dimensionally reduce,
we get
$-{2\ov 3}E^{ij}\nabla_i\nabla_j\phi$ term,
while we get  $2 E^{ij}\nabla_i\nabla_j\phi$ term if we start
directly in 10 dimensions (see  \Lone).
That would be a puzzling contradiction if not
for the fact that the term $E^{ij}\nabla_i\nabla_j\phi$
is, in fact, vanishing on-shell,  thanks to the non-trivial
identity  \ijEij,\kkk\  proved above.
We suspect that there should be several similar identities
related to  supersymmetry of the $R+R^4+...$ actions which
 may explicitly  appear
 in  constructing these actions in the
 component approach  as in \gres.

%%%%%%%%%%%%%%%%%%%%%%%%%%%%%%%%%%%%%%%%%%%%%%%%%%%%%%%%
%%%%%%%%%%%%%%%%%%%%%%%%%%%%%%%%%%%%%%%%%%%%%%%%%%%%%%%%
\vfill\eject
\listrefs
\end

%%%%%%%%%%%%%%%%%%%%%%%%%%%%%%%%%%%%
\appendix{A}{Eguchi-Hanson and KK monopole metrics }
%%%%%%%%%%%%%%%%%%%%%%%%
The Eguchi-Hanson and KK monopole 4-dimensional manifolds can be described
by the metric of the form
\eqn\metreh{
ds_4^2 = \e^{2y}\big( dt^2 +   (d\psi + \cos\theta d\pp )^2 +
\e^{2v}\big(  d\theta^2+ \sin^2\theta d\pp^2\big)\big)\ ,}
where $y$ and $v$ are functions of $t$.
Computing the effective 1-dimensional action for the system, we get
\eqn\lagreh{
S= {1\ov 2}\int dt\
 \e^{2v+2y} \big(  v'^2 +  4v'y'+3y'^2
-{1\ov 4}\e^{-4v}+ \e^{-2v}\big)\ .}
There are two superpotentials for the system
\eqn\Wp{
W_+= \e^{v+2y}\big(  1 + \e^{-v}\big)\ ,}
leading to the following system of first-order equations
\eqn\sysp{
v'= \e^{-v}+\e^{-2v}\ ,\ \ \ \
y'= -{1\ov 2}\e^{-2v}\ ,}
and
\eqn\Wm{
W_-= \e^{v+2y}\big(  1 - \e^{-v}\big)\ ,}
leading to the following system of first-order equations
\eqn\sysm{
v'= \e^{-v}-\e^{-2v}\ ,\ \ \ \
y'= {1\ov 2}\e^{-2v}\ .}
The systems are equivalent and describe the KK monopole. All $R^4$
corrections vanish on the solutions to either \sysp\ or \sysm. Thus the
KK monopole is not corrected in superstring theory.